\documentclass[journal,10pt]{IEEEtran}

\IEEEoverridecommandlockouts 
\usepackage{epsfig,latexsym}
\usepackage{subfigure}
\usepackage{caption}
\usepackage{float}
\usepackage{indentfirst}
\usepackage{amsmath}
\usepackage{amssymb}
\usepackage{cite}
\usepackage{bm}
\usepackage{url}
\usepackage{lastpage}
\usepackage{color}
\usepackage{fancyhdr}
\usepackage{multirow}
\usepackage{setspace}

\usepackage[noend]{algpseudocode}
\usepackage{algorithmicx,algorithm}
\usepackage[square, comma, sort&compress, numbers]{natbib} 
\usepackage{epstopdf}

\normalsize

\begin{document}
\title{Deep Transfer Learning-based Detection for Flash Memory Channels}

\author{
Zhen Mei,~\IEEEmembership{Member,~IEEE},
Kui Cai,~\IEEEmembership{Senior Member,~IEEE},
Long Shi,~\IEEEmembership{Senior Member,~IEEE},
Jun Li,~\IEEEmembership{Senior Member,~IEEE},
Li Chen, ~\IEEEmembership{Senior Member,~IEEE}, and
Kees A. Schouhamer Immink,  ~\IEEEmembership{Life Fellow,~IEEE}
\thanks{This work was supported by the National Natural Science Foundation of
	China under Grants 62201258 and 62071498, by the open research fund of National Mobile Communications Research Laboratory, Southeast University (No. 2023D12), by “the Fundamental Research Funds for the Central Universities”,
	No.30923011035, by the Jiangsu Specially-Appointed Professor Program 2021, by the Singapore Ministry of Education Academic Research Fund Tier 2 T2EP50221-0036 and RIE2020 Advanced Manufacturing and Engineering (AME) programmatic grant A18A6b0057. The material in this paper was presented in part at IEEE International Conference on Communications (ICC), May 2023. {\it{(Corresponding authors: Kui Cai, and Jun Li)}}
	
Zhen Mei is with the School of Electronic and Optical Engineering, Nanjing University of Science and Technology, Nanjing 210000, China. He is also with the National Mobile Communications Research Laboratory, Southeast University, Nanjing 210096, China (email: meizhen@njust.edu.cn).

Kui Cai is with the Science, Mathematics and Technology Cluster, Singapore University of Technology and Design, Singapore 487372 (email: cai\_kui@sutd.edu.sg ).

Long Shi and Jun Li are with the School of Electronic and Optical Engineering, Nanjing University of Science and Technology, Nanjing 210094, China (email: slong1007@gmail.com; jun.li@njust.edu.cn).

Li Chen is with the School of Electronics and Information Technology, Sun Yat-sen University, Guangzhou 510006, China (email: chenli55@mail.sysu.edu.cn ).

Kees A. Schouhamer Immink is with Turing Machines Inc., 3016 Rotterdam, The Netherlands (email: immink@turing-machines.com).

} 
}

\maketitle

\begin{abstract}
The NAND flash memory channel is corrupted by different types of noises, such as the data retention noise and the wear-out noise, which lead to unknown channel offset and make the flash memory channel non-stationary. In the literature, machine learning-based methods have been proposed for data detection for flash memory channels. However, these methods require a large number of training samples and labels to achieve a satisfactory performance, which is  costly. Furthermore, with a large unknown channel offset, it may be impossible to obtain enough correct labels. In this paper, we reformulate the data detection for the flash memory channel as a transfer learning (TL) problem. We then propose a model-based deep TL (DTL) algorithm for flash memory channel detection. It can effectively reduce the training data size from $10^{6}$ samples to less than $10^{4}$ samples. Moreover, we propose an unsupervised domain adaptation (UDA)-based DTL algorithm using moment alignment, which can detect data without any labels. Hence, it is suitable for scenarios where the decoding of error-correcting code fails and no labels can be obtained. Finally, a UDA-based threshold detector is proposed to eliminate the need for a neural network. Both the channel raw error rate analysis and simulation results demonstrate that the proposed DTL-based detection schemes can achieve near-optimal bit error rate (BER) performance with much less training data and/or without using any labels.

\end{abstract}


\section{Introduction}
As a type of emerging non-volatile memories (NVMs), NAND flash memory has been widely applied in various storage systems, ranging from mobile devices to data centers. To increase the storage density, flash memory technologies have been evolved from one bit per cell (single-level-cell (SLC)) to a maximum of four bits per cell (quad-level-cell (QLC)) \cite{cai2017error}. However, the raw bit error rate (RBER) of flash memories with multiple bits per cell becomes larger due to various impairments of the system that are difficult to be predicted and compensated \cite{cai2017error}  before the decoding of error correction codes (ECCs). In particular, the wear-out noise caused by program/erase (P/E) cycling and the data retention noise caused by charge leakage over time dominate the source of errors, and they lead to unknown channel/cell threshold voltage offset and make the flash memory channel non-stationary \cite{cai2012error}.

\begin{figure}[t]
	\centering
	\includegraphics[height=1.2in,width=3.4in]{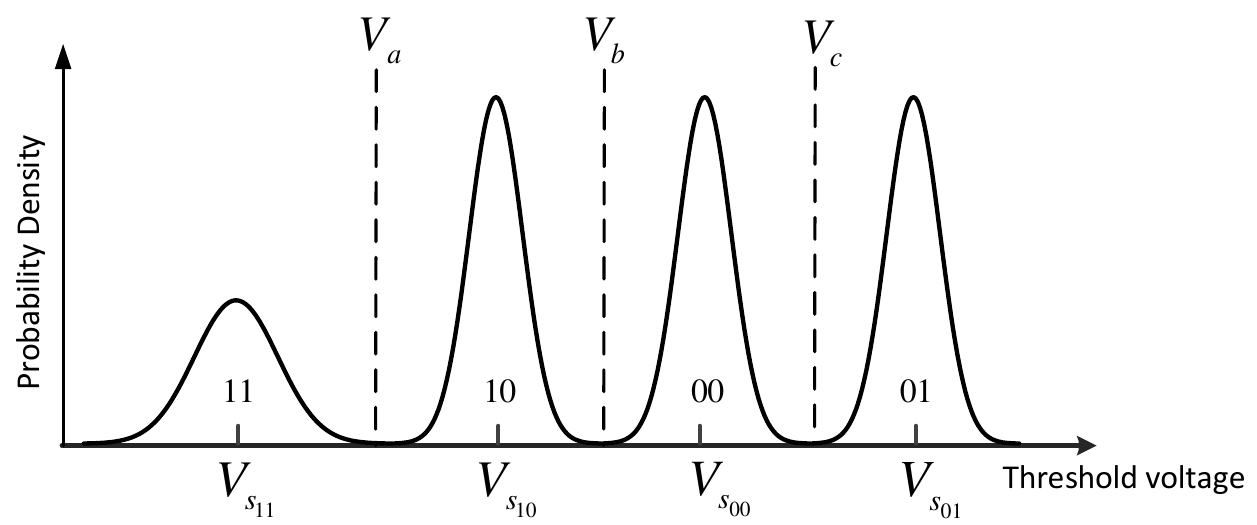}
	\caption{The initial threshold voltage distributions of MLC ($q=2$) NAND flash memory.}
	\label{ini_pdf}
\end{figure}

In this paper, we consider NAND flash memory with $q$ bits stored in each memory cell, which results in $2^{q}$ possible cell states. The threshold voltage of each state can be represented by a probability density function (PDF). As an example, the threshold voltage distributions for multi-level cell (MLC) ($q=2$) NAND flash memory are illustrated in Fig. \ref{ini_pdf}. The boundaries $V_a$, $V_b$, $V_c$ between neighboring states are referred as read reference voltages or read thresholds, which are used to differentiate the states upon reading the memory cell. When the memory cells are corrupted by various noises, the PDF of these states changes, making the original read thresholds no longer optimal. These sub-optimal read thresholds will lead to more raw bit errors, which severely affect the reliability of the flash memory \cite{cai2017error}.

To mitigate such performance degradation of flash memories, ECCs have been employed to correct a certain amount of errors. To correct the multiple-bit errors,  Bose-Chaudhuri-Hocquenghem (BCH) codes have been widely applied in practical flash memories \cite{choi2009vlsi}. As the technology scales down, more powerful ECCs such as the low-density parity-check (LDPC) codes have been proposed to provide higher error-correction capability \cite{zhao2013ldpc, wang2011soft}. Since ECCs correct the errors with the cost of sacrificing data storage efficiency, better channel/data detection schemes that can effectively reduce the channel RBER before ECC decoding are needed. As illustrated in the literature, the most effective approach is to adjust the read thresholds with the change of channel conditions to achieve the best RBER performance, which is also the main focus of this paper.

\subsection{Related Works}
The design of read thresholds for the flash memories has been investigated in many literature works \cite{dong2011use, wang2014enhanced, aslam2016read, mei2019channel, mei2020deep}. They can be classified into two types: the model-driven method and the data-driven method. The model-driven method assumes that the flash memory channel can be modelled by given distributions of the cell threshold voltages, and it then designs the read thresholds based on the estimated PDFs. Specifically, the work of \cite{lee2013estimation} proposed to model the MLC flash memory channel by Gaussian mixture PDF, and then estimated the distribution parameters by gradient descent and Levenberg-Marquardt methods. To better capture the noise statistics, different distributions such as the Beta distribution, Log-normal distribution and Student's t-distribution were also proposed to model the noise PDF \cite{cai2013threshold, luo2016enabling} of the flash memory channel. For given noise distributions, the read thresholds or quantization boundaries can be designed by maximizing the mutual information (MMI) of the channel \cite{wang2014enhanced}, or by optimizing some other information theoretic criteria \cite{aslam2016read, mei2019channel}. However, all these model-based methods assume that the channel modeling and the online parameter estimation are accurate, which is difficult to achieve due to the complication of memory physics for various noises.

On the other hand, the data-driven method does not need to know the {\it{a priori}} noise PDF. One typical approach is to try different read thresholds until the ECC successfully decodes the codeword \cite{cai2015data}. However, this read-retry scheme requires many times of read operations, which lead to a large latency and power consumption. Furthermore, it cannot guarantee to find the optimal read thresholds. Recently, the machine learning-based methods were also proposed to optimize the read thresholds and log-likelihood ratios (LLRs) for ECC decoding \cite{mei2020deep, shi2021convolutional, wiriya2020algorithm, sandell2020machine}. In particular, a deep learning (DL)-based framework to design read thresholds was first proposed in \cite{mei2020deep}, and both recurrent neural network (RNN) and convolutional neural network (CNN)-based detectors were shown to achieve the near-optimal BER performance \cite{mei2020deep, shi2021convolutional}. In \cite{wiriya2020algorithm}, an unsupervised expectation maximization (EM) algorithm was proposed to estimate the channel transition probabilities of the flash memory channel. However, only the binary asymmetric channel (BAC) was actually considered in \cite{wiriya2020algorithm}, and the proposed EM algorithm needs to be iterated with the ECC decoder, leading to a very large latency. In \cite{sandell2020machine}, a machine learning-based LLR estimation scheme was proposed for flash memories. Although it can directly estimate the LLR, it needs to be invoked for each data block and thereby also results in a large latency and power consumption. Moreover, it is difficult to obtain the optimal LLR using training labels in the presence of the unknown channel offset/noise.

\subsection{Motivations and Contributions of This Work}
Although the DL-based read thresholds design can approach the optimal BER performance of the MLC flash memory channel \cite{mei2020deep, shi2021convolutional}, as will be shown in Section III.B, it needs to have a large amount of training samples and labels to achieve a satisfactory performance, which is costly in practice. Even if the neural network can be activated periodically to update the read thresholds, the channel mismatch may still lead to significant performance degradation. Moreover, when the channel offset is large (e.g. caused by a long retention time or a large number of P/E cycles), it will be difficult to obtain enough labels for training. Therefore, it is highly desired to design a DL-based detection approach that can be adapted to unknown channel offset with only a small amount of training samples.

To accomplish this objective, it is necessary to leverage the characteristics of the flash memory channel. It is known that during the early stages of flash memory's lifespan, the channel condition is good and sufficient training samples and labels can be readily acquired. However, as the number of P/E cycles and retention time increase, the noises will severely degrade the performance, and it will become increasingly challenging to obtain labels through ECC decoding. Hence, it is desired to leverage the information from the early stages to reduce the learning difficulty at the targeted P/E cycles and retention time. Inspired by human's capability of learning knowledge from past experience, transfer learning (TL) was proposed to exploit the knowledge and experience gained from a related task to improve the performance of the target task \cite{zhuang2020comprehensive}, which is highly suitable to our application.

There are different types of TL approaches reported in the literature, such as the instance-based TL, the feature-based TL, and the parameter-based TL \cite{pan2009survey, pan2010domain, sun2016return, huang2006correcting, daume2009frustratingly, long2013transfer}. Recently, with the emerge of deep learning techniques, the deep TL (DTL) has also been proposed by integrating TL with deep neural networks \cite{sun2016deep, zhu2020deep}. With TL or DTL, the knowledge from the source domain can be transferred to the target domain under the condition that a connection/similarity exists between the two domains. This will significantly alleviate the training difficulty in the target domain. In particular, as a type of TL, the unsupervised domain adaptation (UDA) techniques such as maximum mean discrepancy (MMD) \cite{borgwardt2006integrating} and correlation alignment (CORAL) \cite{sun2016return} were proposed to transfer knowledge from a labeled source domain to an unlabeled target domain \cite{liu2022deep}.

Inspired by the above ideas, in this work, we propose a DTL framework for the data detection of flash memory channels to reduce the required training samples and labels. Our contributions are summarized as follows.
\begin{enumerate}
\item We formulate the data detection of the flash memory channel as a TL problem. We propose a model-based DTL algorithm which can reduce the number of required training samples and labels by two orders of magnitude.
\item To cope with the situations where labels are difficult to obtain, we propose a UDA-based DTL algorithm by aligning the first-order moments of the source domain and target domain, based on which the neural network can be trained without any labels in the target domain.
\item Inspired by the UDA-based DTL algorithm, we further propose a simple UDA-based threshold detector such that the neural network is not required in both the source and the target domains.
\item We also derive the symbol error rate (SER) and RBER for the uncoded MLC and triple-level-cell (TLC) flash memory channels as the performance benchmark.
\end{enumerate}
Our proposed DTL framework can not only be used for the data detection of the flash memory channels, but also be applied to other data storage or communication channels with the non-stationary nature. The proposed DTL algorithms can be directly applied to data detection with quantized signals. Furthermore, although in this work we adopt the RNN for data detection, the model-based DTL and UDA-based DTL algorithms can be applied to other types of neural networks as well.

The rest of this paper is organized as follows. The basics of NAND flash memories and the corresponding channel model are introduced in Section II. In Section III, we present the RNN-based data detection scheme for flash memories, and the effect of training data size is investigated. In Section IV, we formulate the data detection for flash memories as a TL problem, and we propose two DTL algorithms and a UDA-based threshold detector to effectively reduce the required training samples and labels. Experiment results are illustrated in Section V. Finally,  Section VI concludes the paper.

\section{Preliminaries}

\subsection{NAND Flash Memory Basics}
In NAND flash memory, $q$-bit data is stored as the threshold voltage in each flash memory cell. The possible states of memory cells are denoted as $\left\lbrace  s_{0}, s_{1}, \ldots, s_{2^{q}-1}  \right\rbrace $, where $s_0$ is known as the erased state, and other states are programmed states. For example, for a MLC flash memory, there are four possible states $\left\lbrace  s_{0}, s_{1}, s_{2}, s_{3} \right\rbrace $, and a Gray mapping can be used to represent the bit mapping of each state given by $\left\lbrace {11}, {10}, {00}, {01} \right\rbrace $. Similarly, TLC flash memory has eight possible states $\left\lbrace  s_{0}, s_{1}, \ldots, s_{7}  \right\rbrace $ and the corresponding bit mapping can be taken as $\left\lbrace {{111}}, {{110}}, {{100}}, {{000}}, {{010}},{{011}},{001},{101} \right\rbrace $.

However, the threshold voltage of each state will shift and their distributions overlap due to various types of noises in flash memories. This will result in decision errors and serious degrade the data recovery performance. There are four major sources of errors, namely, programming noise, data retention noise, wear-out noise, and the cell-to-cell interference (CCI) \cite{dong2011use, dong2013enabling}. The characteristics of these four types of noises are described as follows:
\subsubsection{Programming Noise}
Each flash memory cell is a floating gate and its threshold voltage can be configured by transferring charges into the floating gate. However, process variations will lead to the programming noise $n_p$ of each voltage state, which follows a Gaussian distribution with zero mean and variance of $\sigma_e^{2}$ or $\sigma_p^{2}$ \cite{dong2013enabling, wang2014histogram, dong2014using}. Here, $\sigma^{2}_{e}$ denotes the noise variance of the erased state voltage $v_{s_{0}}$, and $\sigma_p^{2}$ represents that of each programmed state voltage $\left\lbrace v_{s_{1}}, v_{s_{2}}, \ldots,  v_{s_{2^{q}-1}} \right\rbrace $. Therefore, we have
	
	\begin{equation}
		p_{n_p}(v) =   
		\begin{cases}
			\mathcal{N}(0, \sigma^{2}_{e}), &\mbox{for $v\in \left\lbrace  v_{s_{0}} \right\rbrace $}\\
			\mathcal{N}(0, \sigma^{2}_{p}), &\mbox{for $v\in \left\lbrace  v_{s_{1} \sim s_{2^{q}-1}} \right\rbrace $}  
		\end{cases}.                                                                                                                                                                                
	\end{equation}
	
Then, this programming procedure is performed by repeatedly pulsing the voltage with a step voltage $\bigtriangleup V_{pp}$, which is known as incremental-step-pulse programming (ISPP) \cite{lee2002effects}. The ISPP noise $n_i$ only affects the programmed states $\left\lbrace v_{s_{1}}, v_{s_{2}}, \ldots,  v_{s_{2^{q}-1}} \right\rbrace $, while the erase state voltage $v_{s_{0}}$ is not affected by $n_i$. It causes the voltage of the programmed state memory cells to follow a uniform distribution \cite{dong2011use}, given by
\begin{equation}
	p_{n_i}(v) =   
	\begin{cases}
		\frac{1}{\bigtriangleup V_{pp}}, &\mbox{$V_p\leq v \leq V_p + \bigtriangleup V_{pp }$}\\
		0, &\mbox{otherwise},  
	\end{cases},                                                                                                                                                                                   
\end{equation}
where $V_p\in \{V_{s_{10}}, V_{s_{00}}, V_{s_{01}}\}$. Hence, the overall distribution of the erase state is $p_{n_p}(v)$, and that of programmed states is the convolution of $p_{n_p}(v)$ and $p_{n_i}(v)$.

\subsubsection{Data Retention Noise}
The data retention noise $n_r$ is caused by charge leakage after the memory cell is being programmed, and it results in a threshold voltage drop over time. Specifically, as the retention time increases, the threshold voltage $v_p$ will shift towards that of the lower-voltage states, while the erased state voltage $v_{s_{0}}$ is almost unchanged. Moreover, the voltage shift of higher-voltage states is larger than that of the lower-voltage states. Following \cite{wang2014histogram, chen2014increasing, dong2012estimating}, the data retention noise can be model by a Gaussian distribution with mean $\mu_{r_{s}}$ and standard deviation $\sigma_{r_{s}}$, given by \cite{dong2013enabling}
\begin{align}
\mu_{r_{s}} &= (V_s - x_0) \cdot ( A_{t}N_{\text{PE}}^{\alpha_{i}} + B_{t}N_{\text{PE}}^{\alpha_{o}} )\cdot \ln(1+T),   \\
\sigma_{r_{s}} &= 0.3|\mu_{r_{s}}|,
\end{align}
where $V_s$ is the desired write voltage level with $s\in \left\lbrace  s_{0}, s_{1}, \ldots, s_{2^{q}-1}  \right\rbrace $, $N_{\text{PE}}$ denotes the number of P/E cycles, $T$ is the retention time, and $(x_0, A_{t}, B_{t}, \alpha_{i}, \alpha_{o})$ are constants.

\subsubsection{Wear-out Noise}
The wear-out noise $n_w$ is caused by the repeated P/E cyclings that damage the oxide layer of floating gate transistors \cite{mielke2004flash}. The wear-out noise tends to widen the threshold voltage distributions and can be modelled by a Gaussian or exponential distribution \cite{wang2014histogram, aslam2018decision} with zero mean and standard deviation of $\sigma_{w}=0.00027 N_{\text{PE}}^{0.62}$ \cite{dong2013enabling}.

\subsubsection{CCI}
Apart from the above noises, the threshold voltage of a flash memory cell may also be affected by the programming of its adjacent cells \cite{lee2002effects}, which is known as the CCI. The CCI happens due to parasitic capacitance coupling between memory cells. The interference from the adjacent cells is linearly added to the threshold voltage of a victim cell. As reported in the literature \cite{dong2010using}, the CCI can be effectively mitigated by pre-distortion or post-processing techniques. Hence, in this work, we assume that the CCI has already been removed.

\subsection{Channel Model}
The overall threshold voltage distributions of flash memory cells can be computed as the convolution integral of all the noise components, and it can be well approximated by the Gaussian distribution \cite{aslam2018decision}.  In this paper, we first adopt the Gaussian model to generate noise samples for simulations. The combined noises can be expressed as
\begin{equation} \label{overall}
 v = V_s + n_i + n_p + n_r + n_w.
\end{equation}
Each noise component in \eqref{overall} follows the Gaussian distribution and the final combined means and variances are given by

\begin{align}
\mu_{s_{0}}&=V_{s_{0}}-\mu_{r_{s_{0}}}, \\
\mu_{s_{1} \sim s_{2^{q}-1}}&=V_{s_{1} \sim s_{2^{q}-1}}+\frac{\bigtriangleup V_{pp}}{2} - \mu_{r_{s_{1} \sim s_{2^{q}-1}}}, \\
\sigma_{s_{0}}^{2}&=\sigma_{e}^{2} + \sigma_{w}^{2} + \sigma_{r_{s_{0}}}^{2},\\
\sigma^{2}_{s_{1} \sim s_{2^{q}-1}}&=\sigma_{p}^{2} + \sigma_{w}^{2} + \sigma_{r_{s_{1} \sim s_{2^{q}-1}}}^{2},
\end{align}
for the erased-state cell and programmed-state cell, respectively. In the simulations, we adopt the parameters from \cite{dong2013enabling} and assume $\bigtriangleup V_{pp}=0.2$, $\sigma_{e} = 0.35$, $\sigma_{p} = 0.05$, $x_0 = 1.4$, $A_t = 0.000035$, $B_t = 0.000235$, $\alpha_i=0.62$, and $\alpha_o=0.3$. 

Note that non-Gaussian distributions such as the Beta distribution and the Gamma distribution can be used to accurately model the heavy-tail of realistic threshold voltage distributions \cite{cai2013threshold}. To further verify our proposed DTL-based detection approaches in more realistic scenarios, the Gamma distribution is also employed in our simulations.

The above channel model is a simplified model of the flash memory channels that is widely adopted by many literature works [11, 12, 36]. Due to the lack of experimental data from the practical flash memories, this channel model is only used to generate data for training and testing the NNs. Note that our transfer learning approaches are not restricted to a specific channel model, and the proposed DTL-based detectors are data-driven and do not require any knowledge of the channel.

\section{Neural Network-based Data Detection without TL}

\subsection{RNN-aided (RNNA) Threshold Detection}

Similar to \cite{mei2020deep}, the data detection of the flash memory channels is formulated as a machine learning problem and the NN is employed to accomplish the task. This problem can be seen as either a classification task or a regression task.  In this work, to efficiently detect multiple data symbols for each NN inference, the data detection is regarded as a regression task.  Specifically, the readback threshold voltage of the $k$-th memory cell is denoted by $v_k$. The input of the NN is given by $\bm{v}=\left\lbrace v_1, v_2, \cdots, v_N \right\rbrace $, where $N$ is the input size of the NN. The outputs of the NN are the estimates $\bm{\tilde{x}}=\left\lbrace \tilde{x}_1, \tilde{x}_2, \cdots, \tilde{x}_N \right\rbrace$ of the labels $\bm{x}$. To conveniently derive the read thresholds and demap labels to binary bits for ECC decoding subsequently, the $2^q$ voltage states $\left\lbrace v_{s_{0}}, \ldots, v_{s_{2^{q}-1}} \right\rbrace $ of the flash memory cell are labeled as $\{ 0, 1 ,\ldots, 2^{q}-1 \}$, respectively.  Therefore, if we denote the set of network parameters as $\bm{\theta}$, the neural network output can be expressed as
\begin{equation}
\bm{\tilde{x}} = f(\bm{v}, \bm{\theta}),
\end{equation}
where $f(\cdot)$ represents the neural network. The stored data in the $k$-th memory cell can be detected by rounding the neural network output $\tilde{x}_k$ to its nearest integer and then demapping it back to binary bits. Our task is to find an neural network model $f(\cdot)$ and the corresponding parameters $\bm{\theta}$ such that the detection error probability is minimized.

\begin{figure}[t]
\centering
\includegraphics[height=2.2in,width=3in]{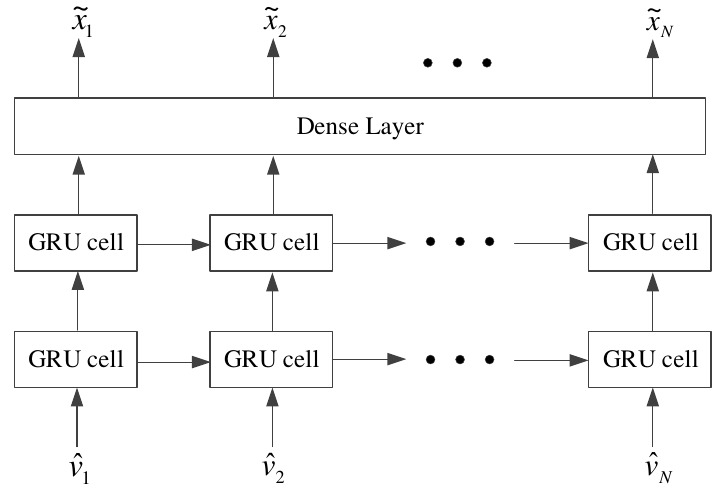}
\caption{The stacked RNN architecture for data detection.}
\label{RNN}
\end{figure}

As illustrated by Fig. \ref{RNN}, we employ the same stacked RNN architecture as proposed in \cite{mei2020deep}. It consists of two gated recurrent unit (GRU) layers and one fully-connected output layer. For the output layer, an additional softplus activation function is used to introduce non-linearity to the neural network, given by $\sigma_{\text{softplus}}(t)=\ln(1+\exp(t))$, with $\sigma_{\text{softplus}}(t)\in [0, \infty)$. Once the RNN architecture is determined, we can train the neural network to find model parameters such that the loss function $\mathcal{L}(\bm{x},\bm{\tilde{x}})$ is minimized. In this work, we choose the mean square error (MSE) as the loss function, given by
\begin{equation}
 \mathcal{L}(\bm{x},\bm{\tilde{x}})=\frac{1}{N}\sum_{k=1}^{N} (x_k-\tilde{x}_k)^2.
\end{equation}
By using the gradient descent-based algorithms and back propagation, the optimized $\bm{\theta}$ can be obtained by minimizing $\mathcal{L}(\bm{x},\bm{\tilde{x}})$ over the entire training data set. After training, we can employ the trained RNN with the optimized $\bm{\theta}$ to detect the data. The corresponding network settings and parameters are given in Table I.

\renewcommand\arraystretch{1.2}
\begin{table}[htbp]
\centering
\caption{Network settings and hyper-parameters.}
\begin{tabular}{|c|c|c|}
\hline
Batch Size  & $20$                      \\ \hline
Number of Epochs  & $50$                      \\ \hline
Loss Function    & MSE                        \\ \hline
Initializer      & Xavier uniform initializer \\ \hline
Optimizer        & Adam optimizer             \\ \hline
\end{tabular}
\end{table}

Similar to \cite{mei2020deep}, we can derive updated read thresholds based on the RNN outputs, leading to the RNN-aided (RNNA) threshold detector. It only needs to be activated periodically when the system is in the idle state. After that, the detection can be carried out directly using the updated thresholds. To obtain labels $\bm{x}$ for training, we can use codewords that are correctly decoded by the ECC decoder as labels. However, if the channel is severely contaminated by noises, it will be difficult to obtain enough labels for training.

\subsection{The Effects of Training Data Size}
As a supervised learning approach, to achieve satisfactory detection performance, a large amount of training samples with the corresponding labels is essential. In \cite{mei2020deep}, it was found that $ 10^{6}$ training samples are sufficient to achieve the near-optimal BER performance. 
However, collecting a large amount of training
data is costly, time-consuming, and consumes additional 
power. A large training data size will also increase the training
complexity. Furthermore, when the channel raw BER is above a certain value, the decoding of ECC will fail. As a result, it may not be feasible to obtain enough correct labels through ECC decoding.

\begin{figure}[t]
\centering
\includegraphics[height=2.4in,width=3.5in]{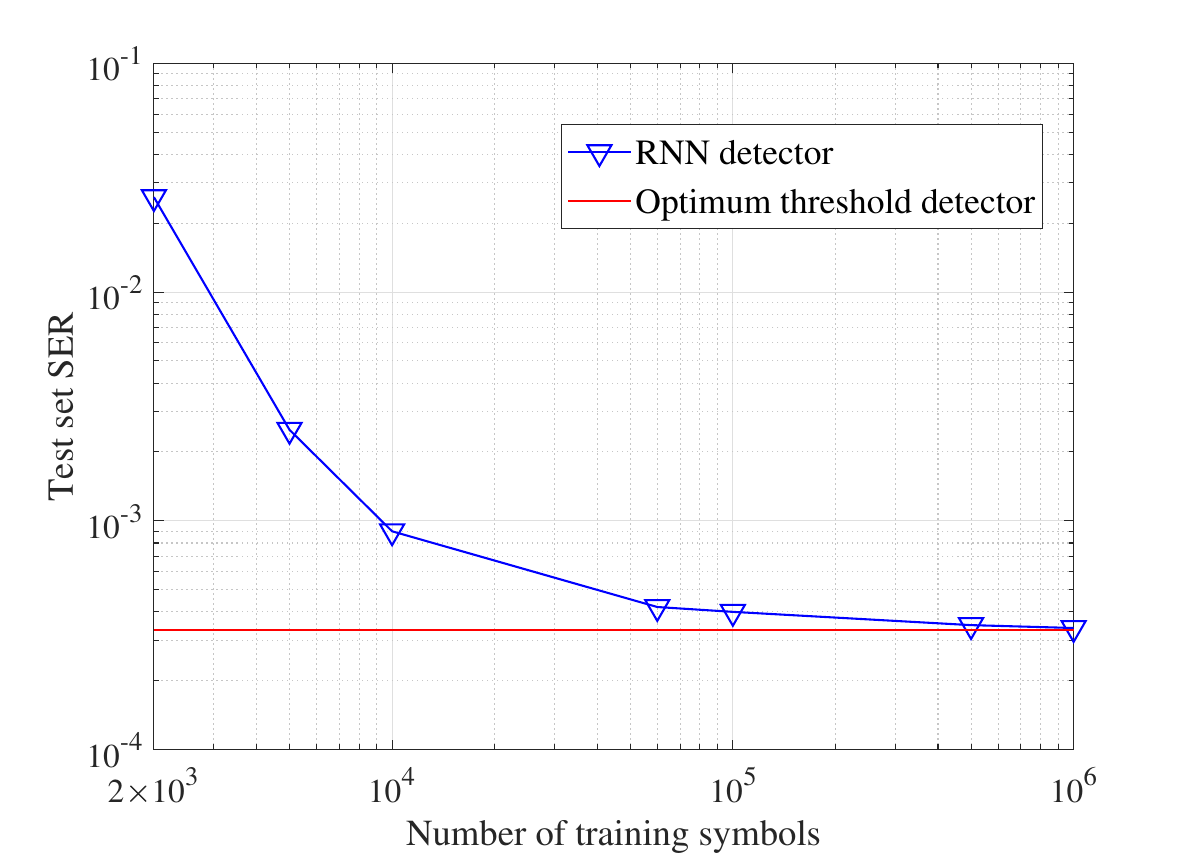}
\caption{RBER performance of the optimum threshold detector and the RNN detector with different number of training samples at $N_{\text{PE}}=10^{3}$ and $T=10^{3}$ hours.}
\label{SER_RNN}
\end{figure}

Fig. \ref{SER_RNN} illustrates the influence of the number of training samples on the RBER performance. The optimal BER which assumes that perfect channel knowledge is known to the detector (derived in Section IV.F) is also included as the benchmark. Observe that the RNN detector can achieve the optimum performance when the number of training samples $N_{\text{train}}=1\times 10^{6}$. When $N_{\text{train}}$ decreases, the BER performance degrades. In the next section, novel DTL-based detection approaches will be presented to reduce the number of training samples and labels without performance degradation.

\section{Transfer Learning-based Data Detection}
In this section, we first formulate the data detection for the flash memory channel as a TL problem, and then propose a model-based TL algorithm to effectively reduce the number of required training samples and labels. We further propose a UDA-based TL algorithm to detect the data without any labels in the target domain. It can work well for the scenarios where the ECC decoding fails due to unknown channel offset/noise.

\subsection{Formulation of TL Problem}
We first define a domain $\mathcal{D}$, which consists of features $\bm{v}$ and labels $\bm{x}$, i.e., $\mathcal{D}=\lbrace {v}_i, x_i  \rbrace_{i=1}^{n}$, where $n$ is the number of samples in the domain. To enable TL, we define two domains, namely, the source domain $\mathcal{D}_s$ and the target domain $\mathcal{D}_t$. Generally, the aim of TL is to transfer the knowledge from the source domain to the target domain, thus improving the performance of the intended task.

In our case, the source domain $\mathcal{D}_s$ consists of $\bm{v}$ and $\bm{x}$ at $N_{\text{PE}}=0$ and $T=0$. It is obvious that we have sufficient training data and labels at the source domain since they are easy to be obtained when the flash memory channel is not severely corrupted by noise in the beginning of its life. The target domain $\mathcal{D}_t$ consists of data and labels at $N_{\text{PE}}=N_{\text{PE}}^{\text{target}}$ and $T=T^{{\text{target}}}$, where $N_{\text{PE}}^{\text{target}}$ and $T^{{\text{target}}}$ are the targeted number of P/E cycles and retention time while performing data detection. Usually, the training samples and labels are limited in the target domain, due to the restriction of read latency, power consumption, and ECC capabilities. However, it is noticed that the source domain and target domain have similar channel characteristics. For example, they have the same number of threshold voltage states, and the statistical distributions of threshold voltages of these states are of the same type (in this work, we follow the literature work and assume the distributions are Gaussian), although the respective values of mean and variance differ. This allows us to apply the DTL technique for the data detection at the target domain with significantly reduced number of training samples and labels.

Some properties of the source domain and target domain for our TL problem are given as follows.
\begin{enumerate}
\item The features of the source domain $\bm{v}_s=\left\lbrace  v_{s, 1}, v_{s, 2}, \ldots, v_{s, n_s} \right\rbrace  $, where $v_{s, k}\in \mathbb{R}$ and $n_s$ is the number of samples in the source domain. The features of the target domain $\bm{v}_t=\left\lbrace  v_{t, 1}, v_{t, 2}, \ldots, v_{t, n_t} \right\rbrace  $, where $v_{t, k}\in \mathbb{R}$, and $n_t$ is the number of samples in the target domain. The feature space of the source domain $\mathcal{V}_s $ is the same as that of the target domain $\mathcal{V}_t$, i.e., $\mathcal{V}_s = \mathcal{V}_t$.

\item The labels of the source domain $\bm{x}_s=\left\lbrace  x_{s, 1}, x_{s, 2}, \ldots, x_{s, n_s} \right\rbrace  $, where $x_{s, k}\in \{ 0, 1 ,\ldots, 2^{q}-1 \}$. The labels of the target domain $\bm{x}_t=\left\lbrace  x_{t, 1}, x_{t, 2}, \ldots, x_{t, n_t} \right\rbrace  $, where $x_{t, k}\in \{ 0, 1 ,\ldots, 2^{q}-1 \}$. The label space of the source domain $\mathcal{X}_s $ is the same as that of the target domain $\mathcal{X}_t$, i.e., $\mathcal{X}_s = \mathcal{X}_t$.

\item The conditional PDF $p(\bm{v}_s|\bm{x}_s)$ of the source domain is different from $p(\bm{v}_t|\bm{x}_t)$ of the target domain, i.e., $p(\bm{v}_s|\bm{x}_s)\neq p(\bm{v}_t|\bm{x}_t)$.
\end{enumerate}
According to above properties, the TL problem in our case can be classified as the homogeneous TL. The task of the TL is to transfer the knowledge from the source domain to the target domain, and learn a prediction function $f:\bm{v}_t \rightarrow \bm{x}_t$ in the target domain to minimize the loss $\mathcal{\phi}$ between the predicted outputs $f(\bm{v}_t)$ and labels $\bm{x}_t$:
\begin{equation}
f^{*} = \mathop{\arg\min}_{f}\mathbb{E}_{(\bm{v}_t, \bm{x}_t)\in \mathcal{D}_t} \mathcal{\phi}(f(\bm{v}_t), \bm{x}_t),
\end{equation}
where $\mathcal{\phi}(f(\bm{v}_t), \bm{x}_t)$  is the loss function between $f(\bm{v}_t)$ and $\bm{x}_t$, which can be defined as the MSE or other loss functions.
In the following, we present DTL methods to accomplish this task based on the RNN detection framework.

\begin{figure}[t]
\centering
\includegraphics[height=1.3in,width=3.4in]{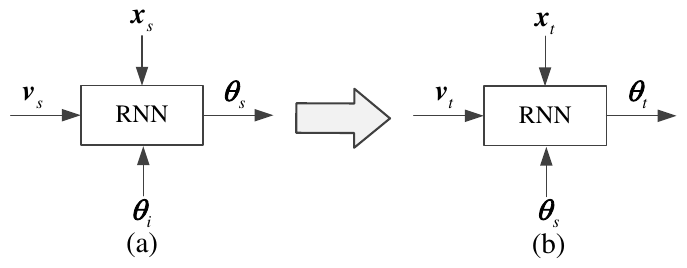}
\caption{Training process of model-based DTL (a) Pre-training (b)  Finetuning (Retraining).}
\label{model_DTL}
\end{figure}

\begin{figure}[t]
\centering
\includegraphics[height=2.7in,width=3.3in]{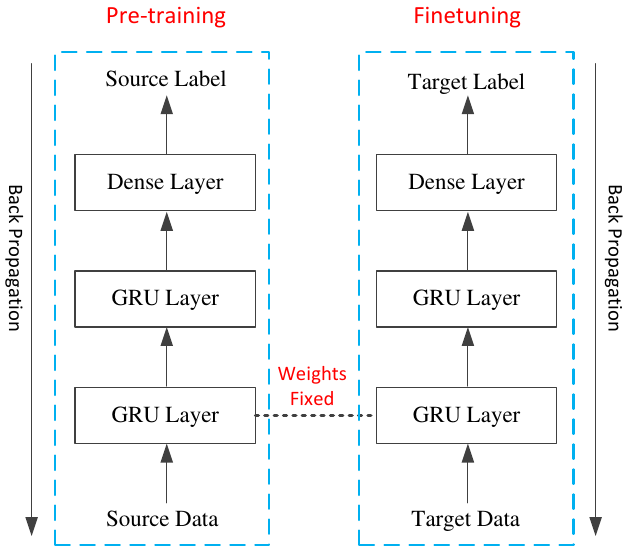}
\caption{Illustration of model-based DTL with weights reuse.}
\label{finuetune}
\end{figure}

\subsection{Model-based DTL}
We first combine DTL directly with the RNN and the resulting model-based DTL is realized by pre-training an RNN in the source domain first, and then transferring the trained RNN model and finetuning the network parameters in the target domain. Specifically, the RNN is trained in the source domain, and the updated set of network parameters $\bm{\theta}_s$ can obtained. Then, we train the same RNN in the target domain with initial parameters $\bm{\theta}_s$. Moreover, to reduce the training complexity, we further propose to freeze some parameters during training.

The proposed model-based DTL is illustrated by Fig. \ref{model_DTL} and Fig. \ref{finuetune}. First, the pre-trained model parameters $\bm{\theta}_s$ are obtained by training the RNN in the source domain with source data $\bm{v}_s$ and labels $\bm{x}_s$. In the target domain, the same RNN architecture is deployed and the model parameters are initialized by $\bm{\theta}_s$. Then, the RNN model is retrained by finetuning the parameters from the pre-trained model. Moreover, it has been demonstrated that in a deep neural network, the first few layers only learn general features and we can directly transfer them to new tasks \cite{yosinski2014transferable}. Motivated by this, as shown in Fig. \ref{finuetune}, we fix model weights of the first GRU layer. Only the second GRU layer and the fully-connected layer are retrained in the target domain, resulting in less training complexity. For example, if the input size of the RNN is 20, the number of training parameters can be reduced from 3921 to 2541, which yields about $35\%$ reduction of the training complexity. The steps of the proposed model-based DTL are summarized in \textbf{Algorithm 1}.

\begin{algorithm}[htbp]
\caption{Model-based DTL detection}\label{alg:1}
\hspace*{0.02in} {\bf Input:}  Source data: $\bm{v}_s$, source labels: $\bm{x}_s$, target data for training: $\bm{v}_t^{\text{train}}$, target labels for training: $\bm{x}_t^{\text{train}}$, target data for testing: $\bm{v}_t^{\text{test}}$.

\vspace*{0.03in}

\hspace*{0.02in} {\bf Output:}  detected symbols in the target domain: $\bm{x}_t$

\begin{algorithmic}[1]
\Statex  \texttt{Training Stage}
\State With $\bm{v}_s$ and $\bm{x}_s$, train the RNN to obtain  model parameters $\bm{\theta}_s$.
\State Initialize the RNN model with $\bm{\theta}_s$, and freeze the weights of the first GRU layer of the RNN model.
\State Finetune the RNN model to obtain parameters $\bm{\theta}_t$ by training the RNN with with $\bm{v}_t$ and $\bm{x}_t$.
\Statex  \texttt{Testing Stage}
\State Detect the target data $\bm{v}_t^{\text{test}}$ using the RNN with $\bm{\theta}_t$ to obtain $\bm{x}_t$.
\end{algorithmic}
\end{algorithm}

\begin{figure}[t]
\centering
\includegraphics[height=2.3in,width=3.5in]{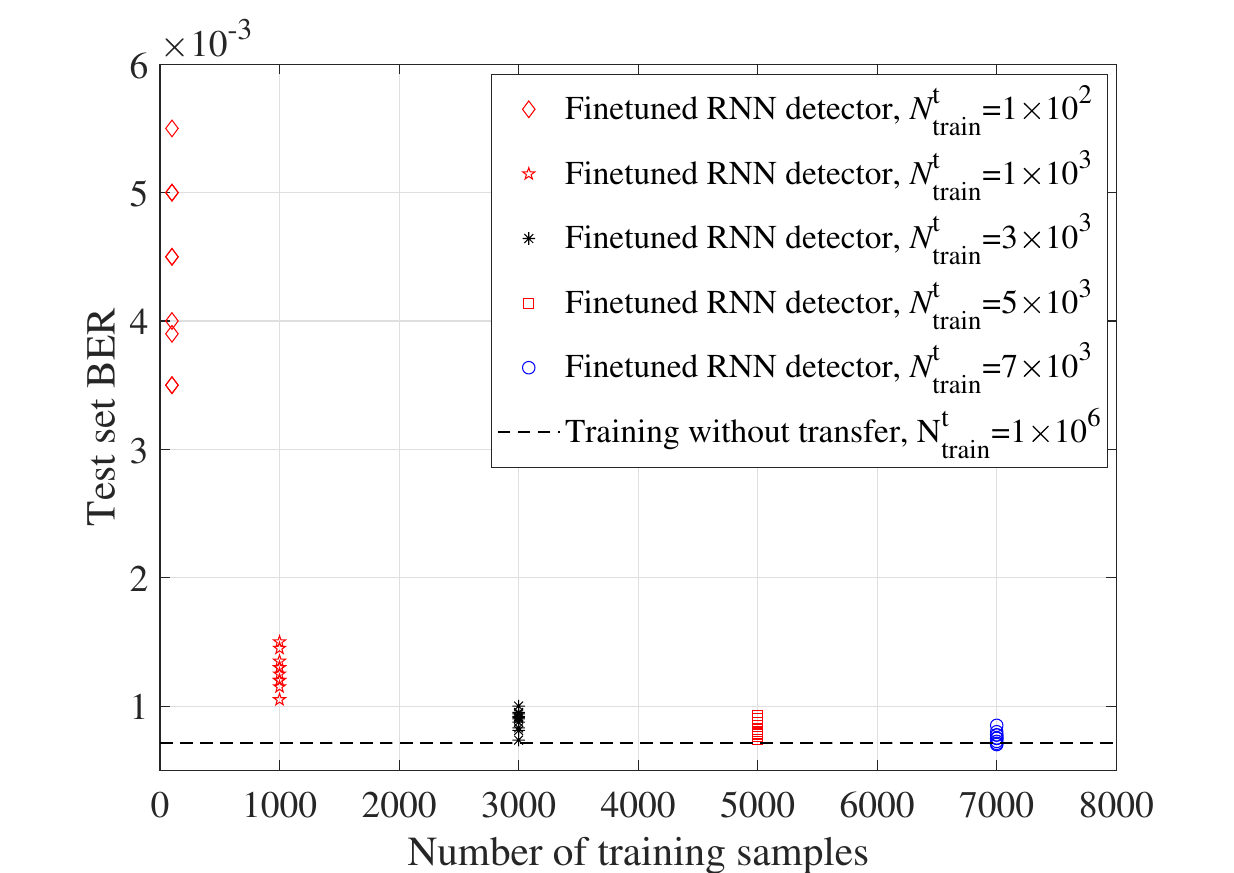}
\caption{The RBER performance of the finetuned RNN detector with different number of training samples in the target domain, with $N_{\text{PE}}^{\text{target}}=5\times 10^{3}$ and $T^{\text{target}}=5\times 10^{3}$ hours.}
\label{SER_finetune}
\end{figure}

To investigate the influence of finetuning on the performance in the target domain, the RBERs of our proposed model-based DTL for MLC flash memory are illustrated by Fig. \ref{SER_finetune}. In the source domain, $1\times 10^{6}$ training samples and labels are used to pre-train the model. In the target domain, we vary the number of training samples $N_{\text{train}}^{\text{t}}$. For each case, to evaluate the stability of our approach, we take 10 trials of RNN finetuning, where the training data are generated randomly for each trial. Observe that as the training data size increases, the BER of the RNN detector becomes more stable, and converges to the optimum. Moreover, with only $7\times 10^{3}$ training samples in the target domain, performance of the proposed model-based DTL can closely approach the best performance where the training with $1\times 10^{6}$ samples is conducted directly in the target domain.

This result indicates that the proposed model-based DTL approach can significantly reduce the training data size by two orders of magnitude. This is due to the main channel characteristics (as described in Section II) of the target domain and the source domain are similar, and thereby it is easier for RNN to learn based on the knowledge also learned from the source domain. Therefore, starting from the pre-trained model parameters, the training process in the target domain can be significantly accelerated.

\subsection{UDA-based DTL}

Although the above described model-based DTL is simple and can achieve excellent performance by finetuning the pre-trained model parameters, it still requires  a certain number of labels in the target domain. However, when the channel offset
is large ({\it e.g.} caused by a long retention time or a large number
of P/E cycles), the decoding of ECC may fail. Then it will be difficult to obtain
enough labels for training the neural network.

As a type of feature-based TL, UDA methods migrate knowledge from a labeled source domain to an unlabeled target domain [29]. Popular UDA approaches include domain alignment with statistic divergence, adversarial learning and so on \cite{liu2022deep}. In this section, we propose a DTL algorithm based on UDA, such that no labels are required in the target domain.  Specifically, we will align the mean of each voltage state between the source domain and the target domain. However, it is difficult to directly calculate the mean of each threshold voltage state in the target domain without any labels due to the overlapping of voltage distributions of different states and the unknown channel offset. To solve this problem, we adopt a $K$-means clustering  approach to find the mean of each threshold voltage state.

The $K$-means clustering algorithm aims to partition the $n$ read-back voltages into $K$ clusters $C=\lbrace  C_0, C_1, \ldots, C_{K-1} \rbrace$ \cite{jain1988algorithms}, such that the intra-cluster distances are minimized. Hence, the optimized clusters $C^{*}$ are given by
\begin{equation} \label{cluster}
C^{*} = \mathop{\arg\min}_{C}\sum_{i=0}^{K-1}\sum_{{v}_j\in C_i}({v}_j-\mu_{t,i})^2,
\end{equation}
where $\mu_{t,i}$ is the mean of the $i$-th cluster in the target domain, and $K=2^q$ since there are $2^q$ voltage states. The $K$-means clustering is an iterative algorithm to find the solution of \eqref{cluster}. At the beginning of the algorithm, the initial centroid of each cluster needs to be determined and it will affect the accuracy and convergence speed of the $K$-means clustering algorithm.
Let the order of centroids at the $k$-th iteration follows
\begin{equation}
\mu_{t,0}^{(k)}<\mu_{t,1}^{(k)} <\cdots<\mu_{t,2^{q}-1}^{(k)}.
 \end{equation}
In our case, it is natural to initialize the centroid of each cluster as
\begin{equation} \label{ini_mean}
\mu_{t,i}^{(0)}=V_{s_{i}}, i=0,1,\ldots, 2^{q}-1,
\end{equation}
since they are the means of initial voltage states. Then, the algorithm proceeds by iterating the following two steps:
\begin{enumerate}
\item \textbf{Assignment step} In the $t$-th iteration, each $v_j$, $j=1,2,\ldots, n$ is assigned to the nearest cluster according to the following equation:
\begin{equation}
i^{*}=\mathop{\arg\min}_{i}(v_j-\mu_{t,i}^{(k)})^2.
\end{equation}
Then, $v_j$ is assigned to cluster $C_{i^{*}}$, $i^{*}\in\lbrace 0,1,\ldots, 2^{q}-1 \rbrace$.
\item \textbf{Updating step} After all $v_j$'s are assigned to the corresponding clusters, the mean of each cluster is updated as
\begin{equation}
\mu_{t,i}^{(k+1)}=\frac{1}{\vert C_i \vert} \sum_{v_j\in C_i} v_j,
\end{equation}
where $\vert C_i \vert$ is the number of elements in $C_i$.
\end{enumerate}
The above iterations are stopped when the mean of each cluster does not change or the maximum number of iterations is reached. By applying this clustering algorithm, we can estimate the mean $\mu_{t, i}$, $i=0,1,\ldots, 2^{q}-1$, of the $i$-th voltage state in the target domain without any labels. 

With $\mu_{t, i}$, we can align the mean of the sub-domain source data as
\begin{equation} \label{mean_align}
\bar{\bm{v}}_{s, i} = \bm{v}_{s, i} - \mu_{s, i} + \mu_{t, i},
\end{equation}
where $\bm{v}_{s, i}$ is the source data that is associated with label $i$, and $\mu_{s, i}$ is the mean of these data. Note that the training data and labels in the source domain are known, and hence $\bm{v}_{s, i}$ and $\mu_{s, i}$ can be easily obtained. With \eqref{mean_align}, the mean of each voltage state in the source domain can be aligned with the target domain. Note that we did not align the variance of the source data with that of the target data as it is difficult to accurately obtain the variance of each voltage state without any labels.

\begin{figure}[t]
\centering
\includegraphics[height=1.2in,width=3.3in]{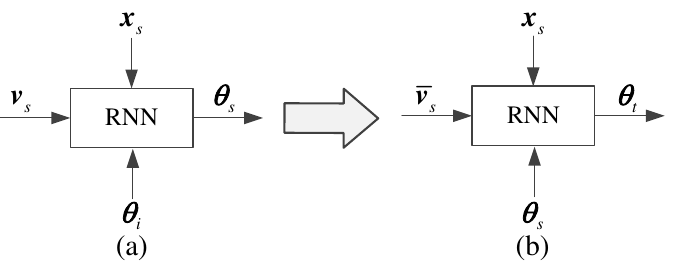}
\caption{Training process of UDA-based DTL (a) Pre-training (b)  Finetuning (Retraining).}
\label{UDA_DTL}
\end{figure}

After the above UDA process, as shown in Fig. \ref{UDA_DTL}, we can finetune the RNN with the transformed source data $\bar{\bm{v}}_{s}$ and the original source labels $\bm{x}_{s}$ to obtain updated model parameters $\bm{\theta}_t$. Moreover, to accelerate the convergence speed and reduce the training complexity, we combine the UDA with the model-based DTL described in Section IV.B. Meanwhile, we can also retrain the pre-trained RNN by fixing some layers and finetuning the remaining parameters. We can then use this updated RNN to detect data in the target domain directly. The detailed procedure of our proposed UDA-based DTL is described in \textbf{Algorithm 2}.

Compared with model-based DTL, the UDA-based DTL does not require any labels in the target domain with the cost of additional DA process. During the UDA process, the most time-consuming step is the $K$-means clustering with a complexity of $O(2^{q}n_t I)$, where $n_t$ is the number of data samples in $\bm{v}_t$, and $I$ is the number of iterations. According to our simulations, with initial centroids given by \eqref{ini_mean}, the proposed $K$-means clustering algorithm for MLC, TLC and QLC can converge with 3, 40, and 150 iterations on average, respectively. Note that although the number of iterations will increase with $q$ increases, the maximum number of $q$ is 4, which is QLC flash memory.  Hence, the number of iterations will not be very large.

\begin{algorithm}[t]
\caption{UDA-based DTL detection}\label{alg:2}
\hspace*{0.02in} {\bf Input:}  Source data: $\bm{v}_s$, source labels: $\bm{x}_s$, target data: $\bm{v}_t$, the initial RNN model.

\vspace*{0.03in}

\hspace*{0.02in} {\bf Output:}  Trained RNN parameters: $\bm{\theta}_t$, detected symbols in the target domain: $\bm{x}_t$

\begin{algorithmic}[1]
\Statex  \texttt{Training Stage}
\State With $\bm{v}_s$ and $\bm{x}_s$, follow model-based DTL to pre-train the RNN to obtain network parameters $\bm{\theta}_s$.
\State Calculate $\bar{\bm{v}}_{s, i}$, $i=0,1,\ldots, 2^{q}-1$ with \eqref{mean_align} to align the sub-domain mean of the source data with that of the target data.
\State Initialize the RNN with $\bm{\theta}_s$, and retrain the RNN with $\bar{\bm{v}}_{s}$ and $\bm{x}_s$ to obtain $\bm{\theta}_t$.
\Statex \texttt{Testing Stage}
\State Detect $\bm{v}_t$ by the trained-RNN with $\bm{\theta}_t$ to obtain $\bm{x}_t$.

\end{algorithmic}
\end{algorithm}

\begin{figure}[t]
\centering
\includegraphics[height=2.3in,width=3.8in]{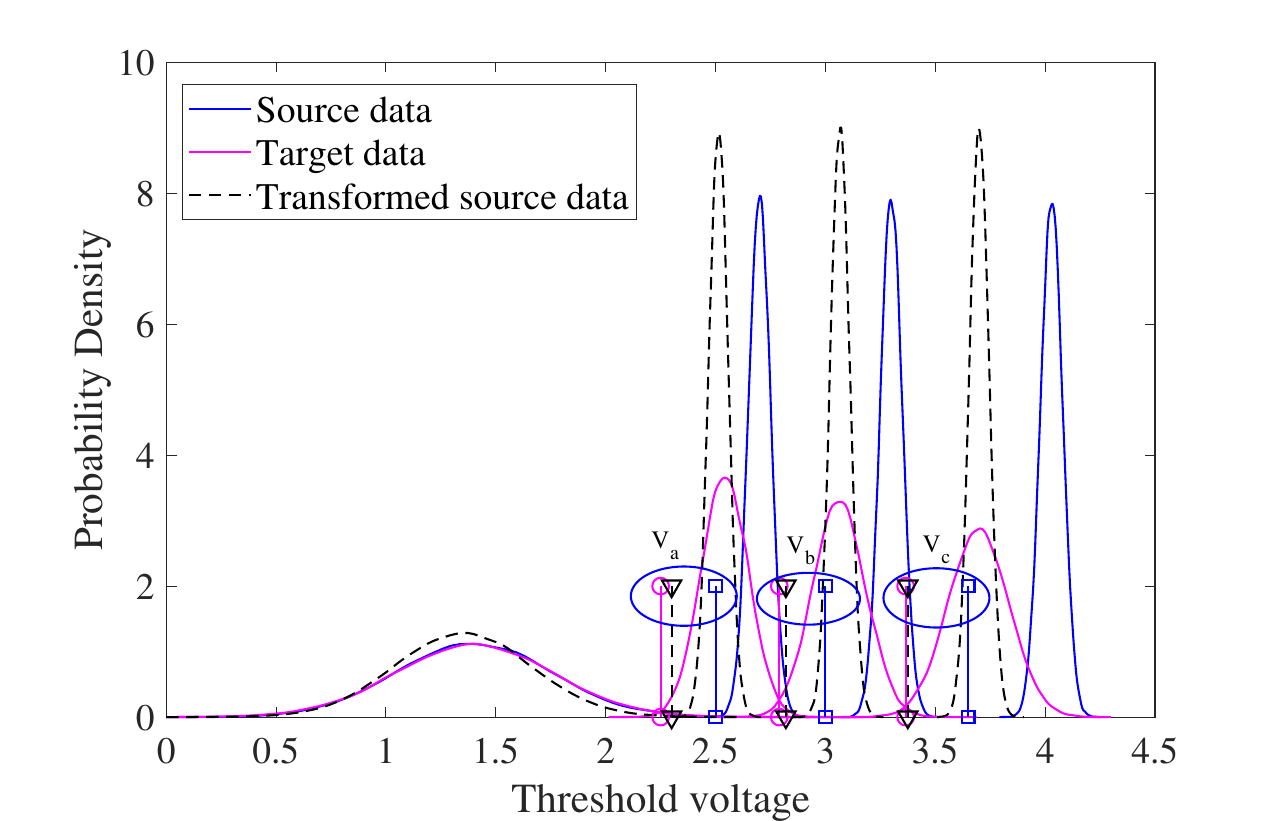}
\caption{The PDFs and learned read thresholds of the source data, target data, and transformed source data for MLC flash memory with $N_{\text{PE}}^{\text{target}}=10^{4}$ and $T^{\text{target}}=10^{4}$ hours (read thresholds markers: square for source data, circle for target data, triangle for transformed source data.) }
\label{Learned_threshold}
\end{figure}

\subsection{RNNA Threshold Detection with DTL}
Although the proposed DTL-based RNN detectors is effective, it is not practical to use the RNN to detect every data block, as it will incur significant read latency and power consumption. To avoid activating the RNN for each data block, we can derive read thresholds from the RNN detected data and then directly use the updated read thresholds for subsequent data detection. We name the corresponding detection RNNA threshold detection with DTL. 

Both the model-based and UDA-based DTL can be used to derive the updated read thresholds for data detection. In particular, for given ${\bm{v}}$ and a set of hard-decision read thresholds $\left\lbrace  V_1^{\text{th}}, V_2^{\text{th}},\ldots,  V_{2^{q}-1}^{\text{th}} \right\rbrace $, we can obtain the detected symbols $\bar{\bm{x}}$. Meanwhile, given ${\bm{v}}$, the RNN can also output its estimated symbols $\tilde{\bm{x}}$. Hence, the RNN learned read thresholds can be obtained by searching for the read thresholds $\left\lbrace  V_1^{\text{opt}}, V_2^{\text{opt}}, \ldots,  V_{2^{q}-1}^{\text{th}} \right\rbrace $ that  minimize the Hamming distance between $\bar{\bm{x}}$ and $\tilde{\bm{x}}$:
 \begin{equation} \label{a123}
\left\lbrace  V_1^{\text{opt}}, V_2^{\text{opt}}, \ldots,  V_{2^{q}-1}^{\text{opt}} \right\rbrace = \mathop{\arg\min}_{\left\lbrace  V_1^{\text{th}}, V_2^{\text{th}},\ldots,  V_{2^{q}-1}^{\text{th}} \right\rbrace } d(\bar{\bm{x}}, \tilde{\bm{x}}).
\end{equation}
To obtain $\left\lbrace  V_1^{\text{opt}}, V_2^{\text{opt}}, \ldots,  V_{2^{q}-1}^{\text{opt}} \right\rbrace$ efficiently, we first uniformly quantize the search space into $m$ intervals (typical value of $m\geq 100$), with boundaries $b_0, b_1, \ldots, b_m$, where $b_0=-\infty< b_1 < \cdots < b_{m-1} < b_m=\infty$. Then, the problem becomes finding $2^{q}-1$ thresholds from $b_0, b_1, \ldots, b_m$ , such that the Hamming distance between $\bar{\bm{x}}$ and $\tilde{\bm{x}}$ is minimized. For MLC flash memory, $2^{q}-1=3$ and for TLC flash memory, $2^{q}-1=7$. To solve this problem, we can adopt a dynamic programming (DP) approach with complexity $\mathcal{O}((2^{q}-1)m^2)$, and the details can be found in \cite{mei2020deep}. After obtaining the updated read thresholds, we can use them directly for data detection. As will be shown in Section V, the read thresholds learned from DTL can achieve near-optimal detection BER performance.

Fig. \ref{Learned_threshold} shows the PDFs of the source data, target data, and transformed source data, with the associated read thresholds, with $N_{\text{PE}}^{\text{test}}=10^{4}$ and $T^{\text{test}}=10^{4}$ for the MLC flash memory. The read thresholds of the target data and transformed source data are derived according to \eqref{a123}. Observe that although the PDF of the transformed source data is more similar to that of the source data than the target data, the read thresholds learned from the transformed source data and the target data are quite close. This indicates that the RNN trained by the transformed source data can achieve a good detection performance. On the other hand, the read thresholds based on the source data are far from that based on the target data, and hence will lead to a severe performance degradation.

\subsection{UDA-based Threshold Detection with Original Read Thresholds}
Inspired by the threshold detector with the UDA-based DTL described above, we further propose a simple UDA-based threshold detection scheme, for which the neural network is not needed in the target domain and we can directly use the original read thresholds in the source domain to perform detection.

That is, by employing the $K$-means clustering algorithm described in Section IV.C in the target domain, we can estimate the mean $\mu_{t, i}$ of the $i$-th voltage state. However, instead of updating the source data in the UDA-based DTL, we update the target data, given by
 \begin{equation} \label{mean_align_inverse}
\bar{\bm{v}}_{t, i} = \tilde{\bm{v}}_{t, i} - \mu_{t, i} + \mu_{s, i},
\end{equation}
where $\tilde{\bm{v}}_{t, i}$ can be obtained by using the pseudo labels obtained from the $K$-means clustering. With \eqref{mean_align_inverse}, the mean of each voltage state in the target domain is aligned with that in the source domain. Then, these target data can be detected by using the optimal read thresholds $\left\lbrace V_1^{s}, V_2^{s}, \dots, V_{2^{q}-1}^{\text{s}} \right\rbrace $ in the source domain.

In this way, the RNN in the target domain is no longer needed for the data detection. Moreover, in practical flash memories, the initial read thresholds $\left\lbrace V_1^{s}, V_2^{s}, \dots, V_{2^{q}-1}^{\text{s}} \right\rbrace $ have already been provided by manufacturers. Therefore, the RNN in the source domain is not needed as well. The details of our proposed UDA-based threshold detection is given in \textbf{Algorithm 3}.

\begin{algorithm}[t]
\caption{UDA-based threshold detection}\label{alg:3}
\hspace*{0.02in} {\bf Input:}  Source data: $\bm{v}_s$, source labels: $\bm{x}_s$, source domain read thresholds: $ V_1^{s}, V_2^{s}, \dots, V_{2^{q}-1}^{\text{s}} $, target data: $\bm{v}_t$.

\vspace*{0.03in}

\hspace*{0.02in} {\bf Output:}  Detected symbols in the target domain: $\bm{x}_t$

\begin{algorithmic}[1]
\State With $\bm{v}_s$ and $\bm{x}_s$, calculate the mean  of each voltage state for the source data, given by $\mu_{s, i}, i=0,1,\ldots, 2^q-1$.
\State With $\bm{v}_t$, find the mean of each voltage state for the target data using $K$-means clustering algorithm in Section IV.C.
\State Calculate $\bar{\bm{v}}_{t, i}$, $i=0,1,\ldots, 2^q-1$ with \eqref{mean_align_inverse} to align the sub-domain mean of the target data with that of the source data.
\State Detect the transformed target data $\bar{\bm{v}}_{t}$ with read thresholds $\left\lbrace V_1^{s}, V_2^{s}, \dots, V_{2^{q}-1}^{\text{s}} \right\rbrace $ to obtain $\bm{x}_t$.

\end{algorithmic}
\end{algorithm}

\subsection{Uncoded Error Rate Analysis}
In this subsection, we derive the the optimum uncoded SER and BER by assuming the channel knowledge is perfectly known. They serve as benchmarks of the various detectors proposed earlier. For given hard-decision read thresholds $\left\lbrace  V_1^{\text{th}}, V_2^{\text{th}}, \ldots,  V_{2^{q}-1}^{\text{th}} \right\rbrace $, assuming that the memory cells are programmed into $2^q$ voltage states with equal probabilities, the SER is calculated as
\begin{align} \nonumber
P_{s} &= \sum_{i=0}^{2^{q}-1} P(v_{s_{i}})P(e|v_{s_{i}})  \\  \nonumber
&= \frac{1}{2^q}\bigg(  P(v>V_1^{\text{th}} | v_{s_{0}}) + \sum_{i=1}^{2^{q}-2}P(v<V_i^{\text{th}} \cup v>V_{i+1}^{\text{th}}  | v_{s_{i}}) \\
& + P(v<V_{2^{q}-1}^{\text{th}} | v_{s_{2^{q}-1}}) \bigg),
\end{align}
where $e$ denotes the event that an error occurs. With given PDF $p_{v_{s_{i}}}, i=0,1,\ldots, 2^{q}-1$ for each voltage state $v_{s_{i}}$, the final expression of the SER is given by
\begin{align} \nonumber
P_{s} &= \frac{1}{2^q}\bigg( \int_{V_1^{\text{th}}}^{\infty}p_{v_{s_{0}}}dv + \sum_{i=1}^{2^{q}-2} \bigg( \int_{-\infty}^{V_i^{\text{th}}}p_{v_{s_{i}}}dv + \\ \label{SEP}
&  \int_{V_{i+1}^{\text{th}}}^{\infty}p_{v_{s_{i}}}dv  \bigg) + \int_{-\infty}^{V_{2^{q}-1}^{\text{th}}}p_{v_{s_{2^{q}-1}}}dv \bigg).
\end{align}
By searching read thresholds $\left\lbrace  V_1^{\text{th}}, V_2^{\text{th}},\ldots,  V_{2^{q}-1}^{\text{th}} \right\rbrace $ that minimize \eqref{SEP}, we can obtain the optimum SER, which can serve as the lower bound of the proposed detectors. Next, the corresponding BER can be estimated from the SER. Note that when the Gray mapping is adopted, the adjacent voltage states only differ by one bit. Hence, assuming that only the adjacent states dominate the errors, the BER can be estimated as
\begin{equation}\label{BEP}
P_b \approx P_s/q.
\end{equation}
Note that \eqref{BEP} is a lower bound of the exact BER, since it only considers errors occurred in the adjacent states. However, for the case of TLC where intervals between adjacent voltage levels are getting smaller than the MLC case, more states will affect the BER. Hence, \eqref{BEP} is not accurate enough. To obtain a more accurate BER estimation, we consider adjacent states that differ by either 1-bit or 2-bit. Therefore, the BER is given by
 \begin{equation}\label{BEP_2}
P_b \approx \sum_{i=0}^{2^{q}-1} P(v_{s_{i}})\left( \frac{1}{q}P(e_1|v_{s_{i}})+  \frac{2}{q} P(e_2|v_{s_{i}})\right) ,
\end{equation}
where $e_1$ denotes the event that an error occurs in the adjacent states, while $e_2$ represents the event that an error occurs in the non-adjacent states which results in 2-bit errors per symbol. In \eqref{BEP_2}, when $i=0$, $P(e_1|v_{s_{i}})$ and $P(e_2|v_{s_{i}})$ are given by
\begin{align} \nonumber
P(e_1|v_{s_{i}}) &= P(V_1^{\text{th}}<v< V_2^{\text{th}}| v_{s_{0}}) \\
& = \int_{V_1^{\text{th}}}^{V_2^{\text{th}}}p_{v_{s_{0}}}dv
\end{align}
and
\begin{align} \nonumber
P(e_2|v_{s_{i}}) &= P(v>V_2^{\text{th}}| v_{s_{0}}) \\
& = \int_{V_2^{\text{th}}}^{\infty}p_{v_{s_{0}}}dv.
\end{align}
Similarly, for $i=1,2,\ldots, 2^q-1$, $P(e_1|v_{s_{i}})$ and $P(e_2|v_{s_{i}})$ can be calculated. By substituting $P(e_1|v_{s_{i}})$ and $P(e_2|v_{s_{i}})$ into \eqref{BEP_2}, a more accurate BER estimation can be obtained.

\begin{figure}[t]
\centering
\includegraphics[height=2.6in,width=3.6in]{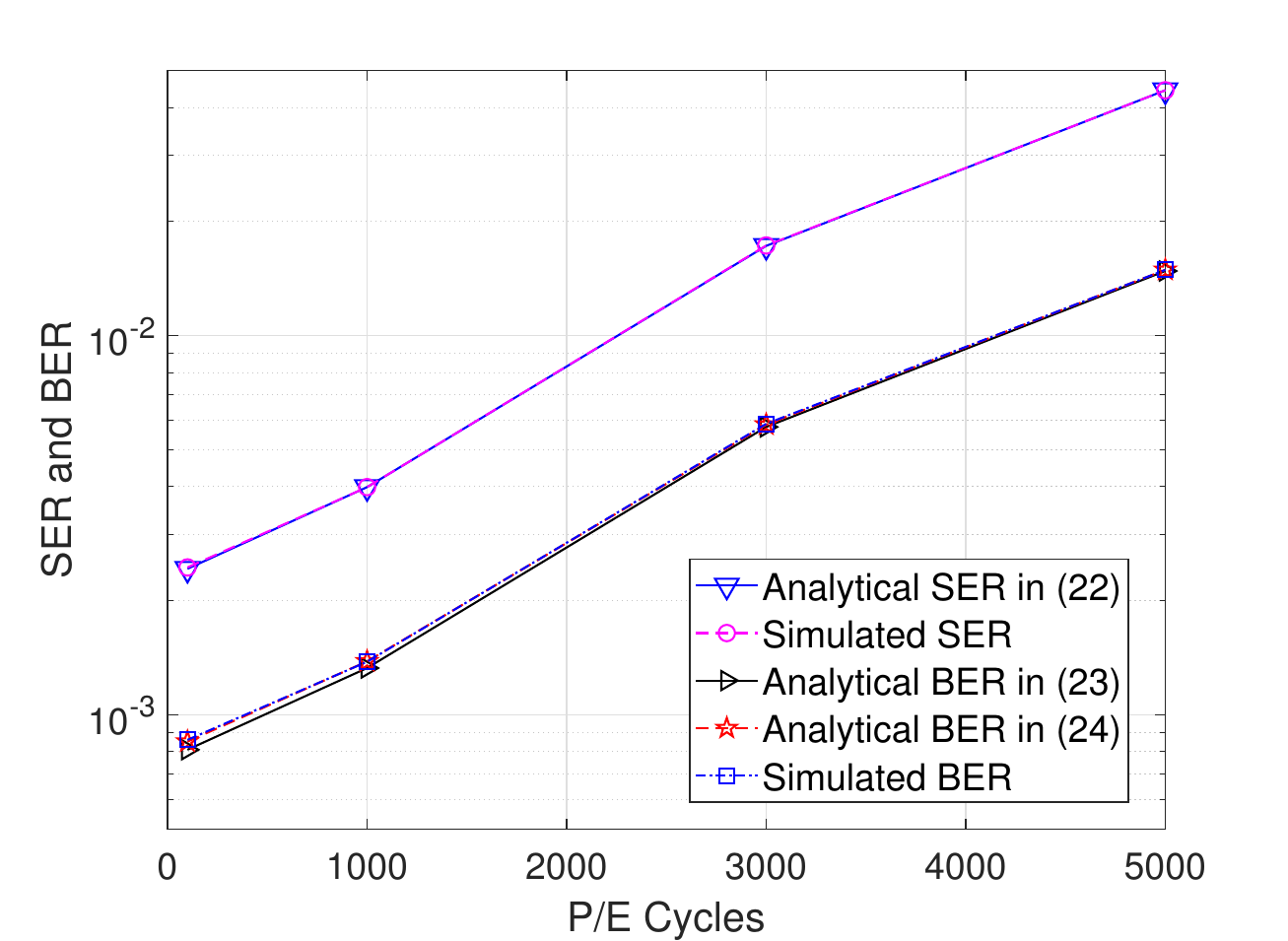}
\caption{The analytical and simulated SER and RBER performance of TLC flash memory at retention time $1\times 10^{4}$ hours.}
\label{SER_BER}
\end{figure}

As illustrated by Fig. \ref{SER_BER},  \eqref{SEP} and \eqref{BEP_2} can match well with the simulated SERs and BERs for the TLC case, while \eqref{BEP} slightly underestimate the BER when the number of P/E cycles is less than 1000.  Therefore, \eqref{SEP} and \eqref{BEP_2} can serve as references for the SER and BER of the proposed detectors, respectively.

\renewcommand\arraystretch{1.5}
\begin{table*}[t]
	\centering
	\caption{The complexity of DTL-based detectors for flash memories}
\begin{tabular}{|c|c|c|c|l}
	\cline{1-4}
	& Model-based DTL detector & UDA-based DTL detector & UDA-based threshold detector &  \\ \cline{1-4}
	RNN Training               &         $\mathcal{O}(\frac{n_{t}}{N} SL(D+L))$                 &              $\mathcal{O}(\frac{n_{t}}{N} SL(D+L))$          &                 0             &  \\ \cline{1-4}
	$K$-means Clustering       &         0              &            $\mathcal{O}(2^{q}n_t I)$            &               $\mathcal{O}(2^{q}n_t I)$                  &  \\ \cline{1-4}
	Read Thresholds Derivation &     $\mathcal{O}((2^{q}-1)m^2)$                     &          $\mathcal{O}((2^{q}-1)m^2)$               &                  0            &  \\ \cline{1-4}
\end{tabular}
\end{table*}

\subsection{Discussions}

\subsubsection{ Comparison of Proposed Detectors}
Comparing the proposed three detection schemes, the RNNA threshold detection with model-based DTL (presented in Section IV.D) is supervised, while the RNNA threshold detection with UDA-based DTL (presented in Section IV.D) as well as the UDA-based threshold detection (presented in Section IV.E) are unsupervised. Therefore, these two detection schemes based on UDA are more suitable for severe channel conditions where labels are not available. On the other hand, compared with the RNNA threshold detection with model-based DTL, these detection schemes require additional DA steps to approach the performance with the model-based DTL.

Unlike the RNNA threshold detection with either the model-based DTL or the UDA-based DTL, the UDA-based threshold detection does not require the neural network in both the source domain and the target domain. Therefore, it is simple to implement. However, it needs to update each target domain data using \eqref{mean_align_inverse} before detection, which will incur additional latency. On the other hand, the RNNA threshold detection with the model-based DTL or the UDA-based DTL can directly use the updated read thresholds to detect data. Therefore, the proposed three knowledge transfer-based detection schemes should be applied according to the availability of labels and system requirements. 

\subsubsection{Complexity Analysis}

In this part, the complexity of our proposed DTL-based detectors is analyzed. First, the proposed RNN architecture consists of two GRU layers and one output layer. For the GRU layer with $L$ GRU cells and input dimension $D$, the number of parameters is $3L(D+L+1)$. In our proposed RNN architecture, $N$ readback threshold voltages are fed into the network for each training or inference. Hence, given a total of $n_{t}$ training samples in the target domain and $S$ training epochs, the training complexity can be approximated by $O(\frac{n_{t}}{N} SL(D+L))$. The inference complexity can be estimated by reducing the number of epochs to one. In this work, $L=20$ and $N=20$. We have $D=1, 20$ for the first and second GRU layers, respectively. Second, the complexity of $K$-means clustering for the UDA-based DTL detection and UDA-based threshold detection is given in Section IV.C. Third, the complexity for deriving read thresholds is given in Section IV.D, which is $\mathcal{O}((2^{q}-1)m^2)$. To obtain good performance, the typical value of $m$ is 100 or larger. Hence, the overall complexity for different detection schemes in the target domain is summarized in Table II.

First, it can be seen from Table II that the training complexity is linearly proportional to the number of training samples. Since it has been demonstrated in Section IV.B that compared with the DL-based approach [10], the proposed DTL algorithms can reduce the number of training samples and labels in the target domain by two orders of magnitude, the training complexity is also decreased by two orders of magnitude. Second, as described in Section IV.C, the maximum value of $q$ is 4, thereby we conclude that the overall complexity will be dominated by the RNN training.

\section{Numerical and Simulation Results}

In this work, the RNN-based DTL algorithms are implemented with Pytorch DL framework \cite{paszke2019pytorch}. In the following, the channel RBER and the LDPC-coded BER for MLC and TLC flash memory channels are evaluated with our proposed threshold detectors. In our experiments, the number of training samples and labels in the target domain is $10^{6}$ for the RNNA threshold detector trained without TL (presented in Section III.B), while it is taken to be $10^{4}$ for the RNNA threshold detectors with DTL. In particular, for the RNNA threshold detector with UDA-based DTL, labels in the target domain are not needed. We evaluate both the uncoded RBER and the LDPC-coded BER performance of various threshold detectors, for both the MLC and TLC flash memories. Note that for the MLC case, we take $V_{s_{0}\sim s_{3}} = \left\lbrace 1.4, 2.6, 3.2, 3.93 \right\rbrace$, while for the TLC case, we set $V_{s_{0}\sim s_{7}} = \left\lbrace 1.4, 2.2, 2.6, 3.0, 3.4, 3.8, 4.2, 4.6 \right\rbrace$.

\begin{figure}[b]
\centering
\includegraphics[height=2.5in,width=3.5in]{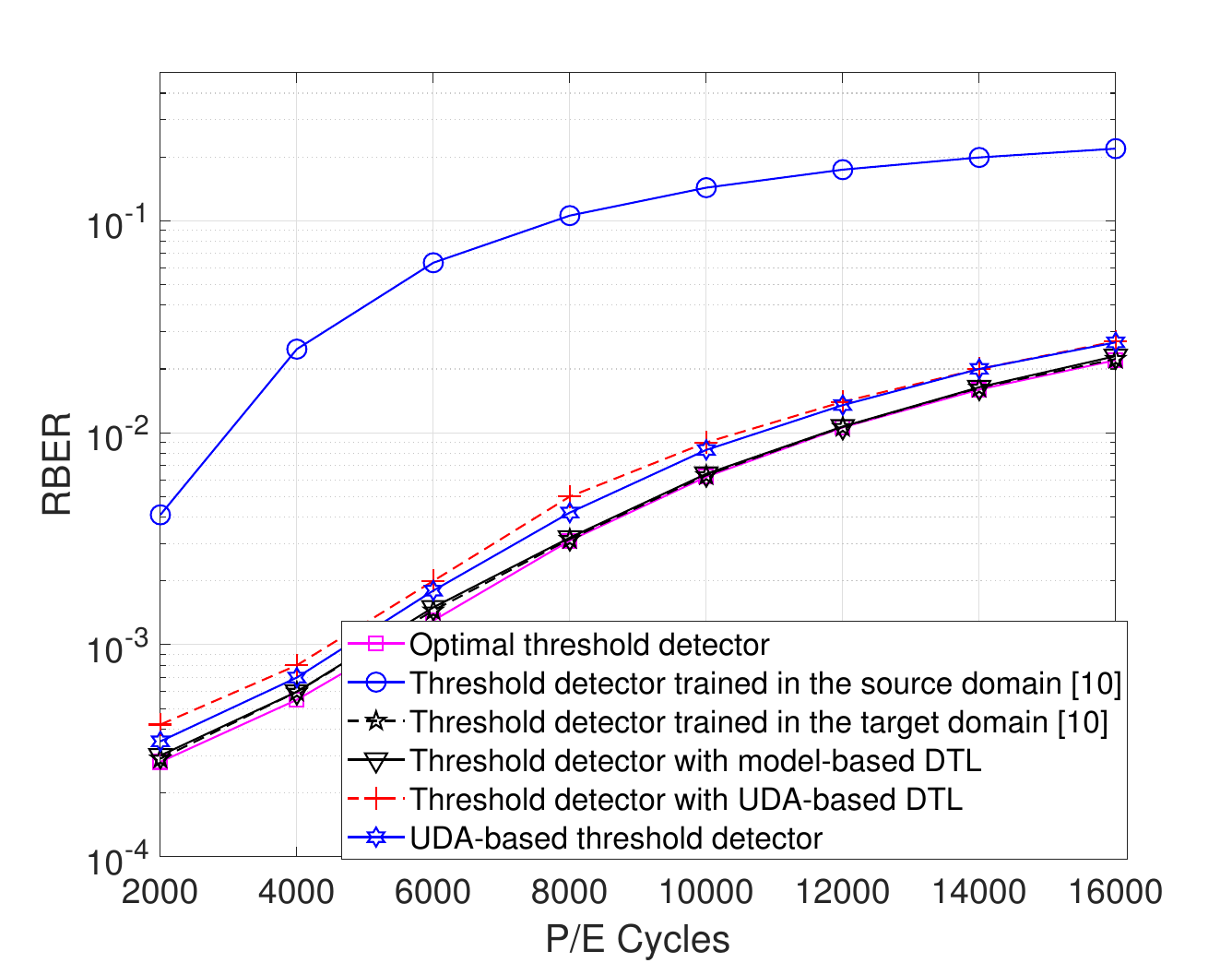}
\caption{The RBER performance of different threshold detectors for MLC flash memory at $T^{\text{target}}=1.2\times 10^{4}$ hours.}
\label{BER_PE_Gaussian}
\end{figure}

\begin{figure}[t]
\centering
\includegraphics[height=2.5in,width=3.5in]{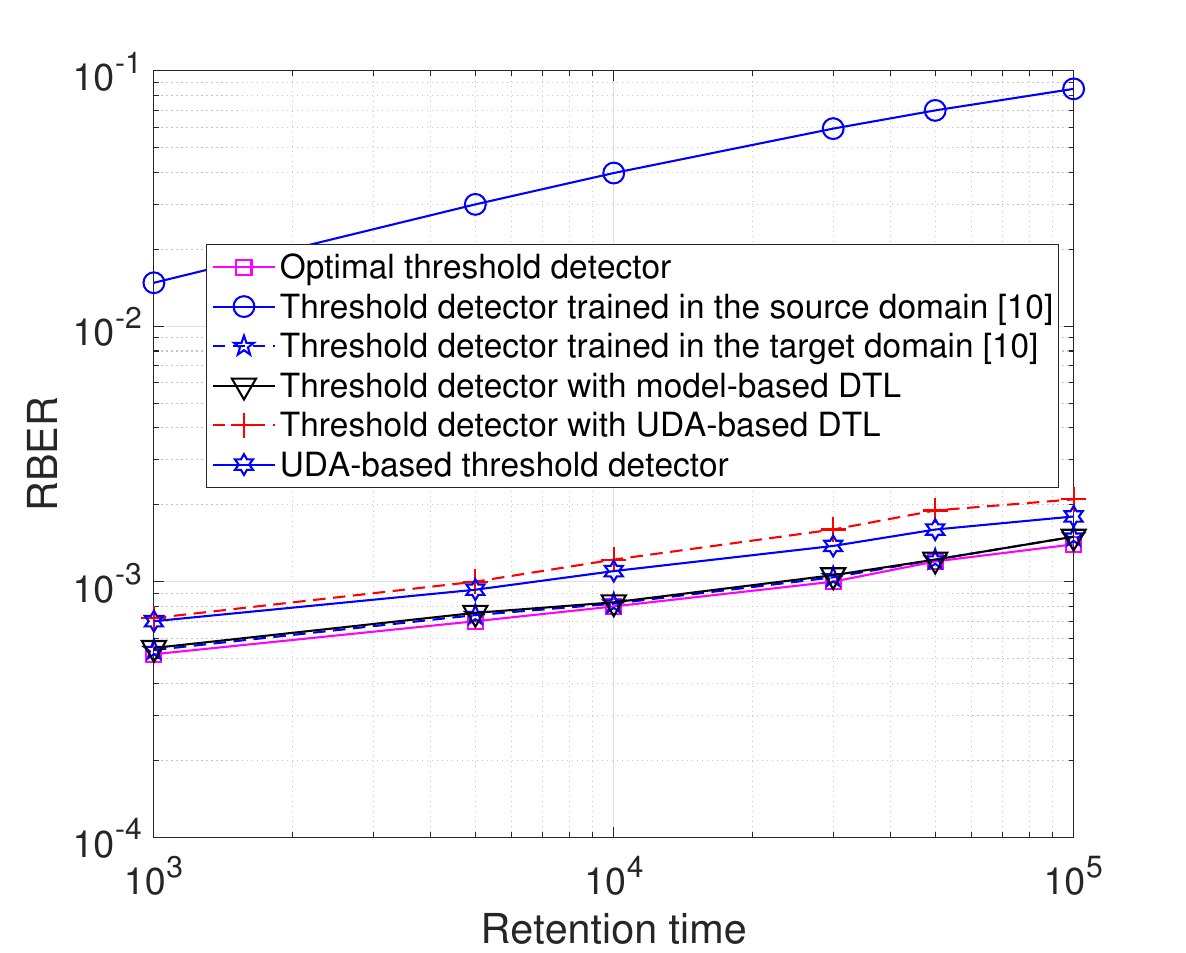}
\caption{The RBER performance of different threshold detectors for MLC flash memory at $N_{\text{PE}}^{\text{target}}=5\times 10^{3}$.}
\label{BER_time_Gaussian}
\end{figure}

\begin{figure}[t]
\centering
\includegraphics[height=2.5in,width=3.5in]{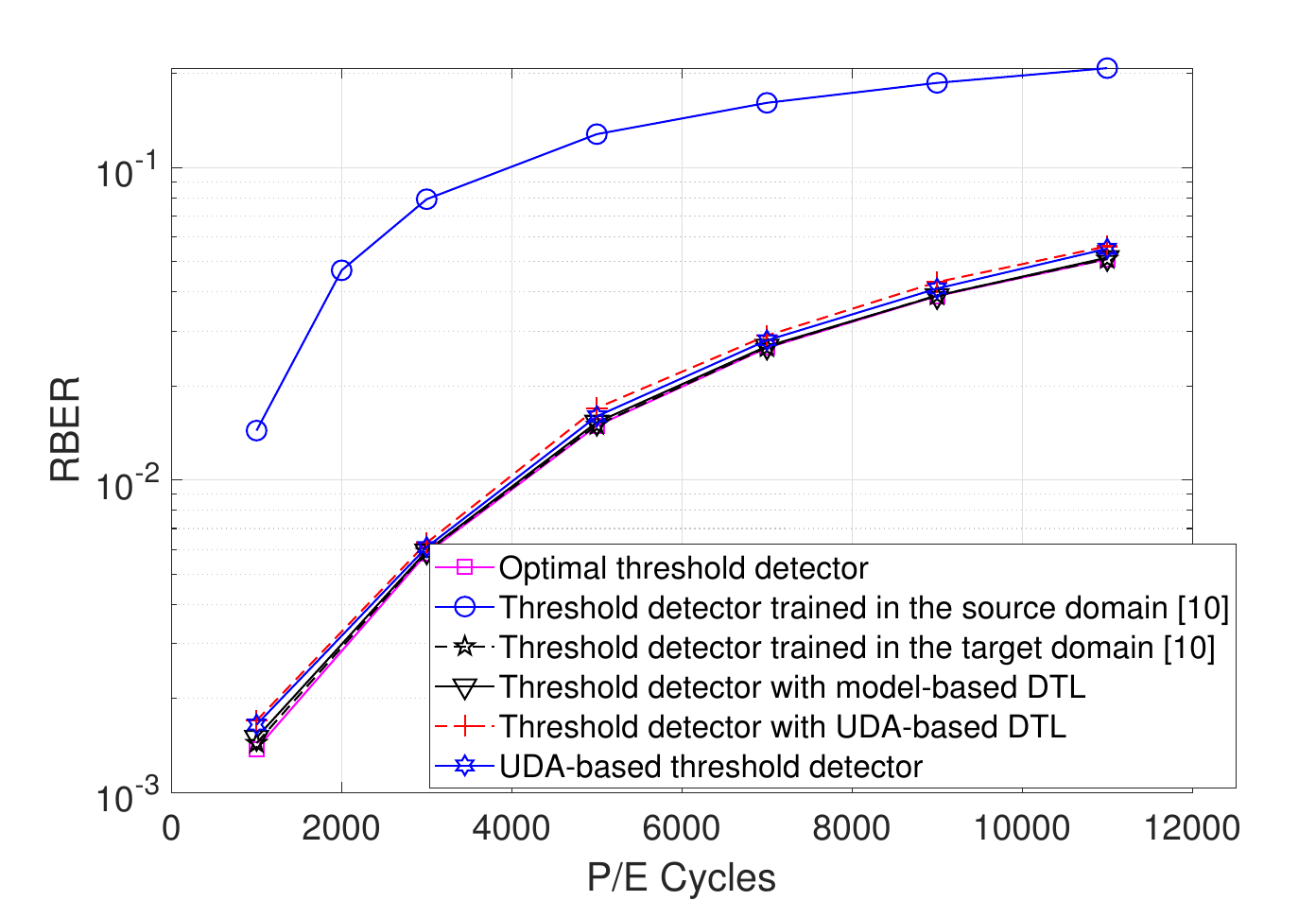}
\caption{The RBER performance of different threshold detectors for TLC flash memory at  $T^{\text{target}}=1\times 10^{4}$ hours.}
\label{BER_PE_Gaussian_TLC}
\end{figure}

\begin{figure}[t]
	\centering
	\includegraphics[height=2.5in,width=3.5in]{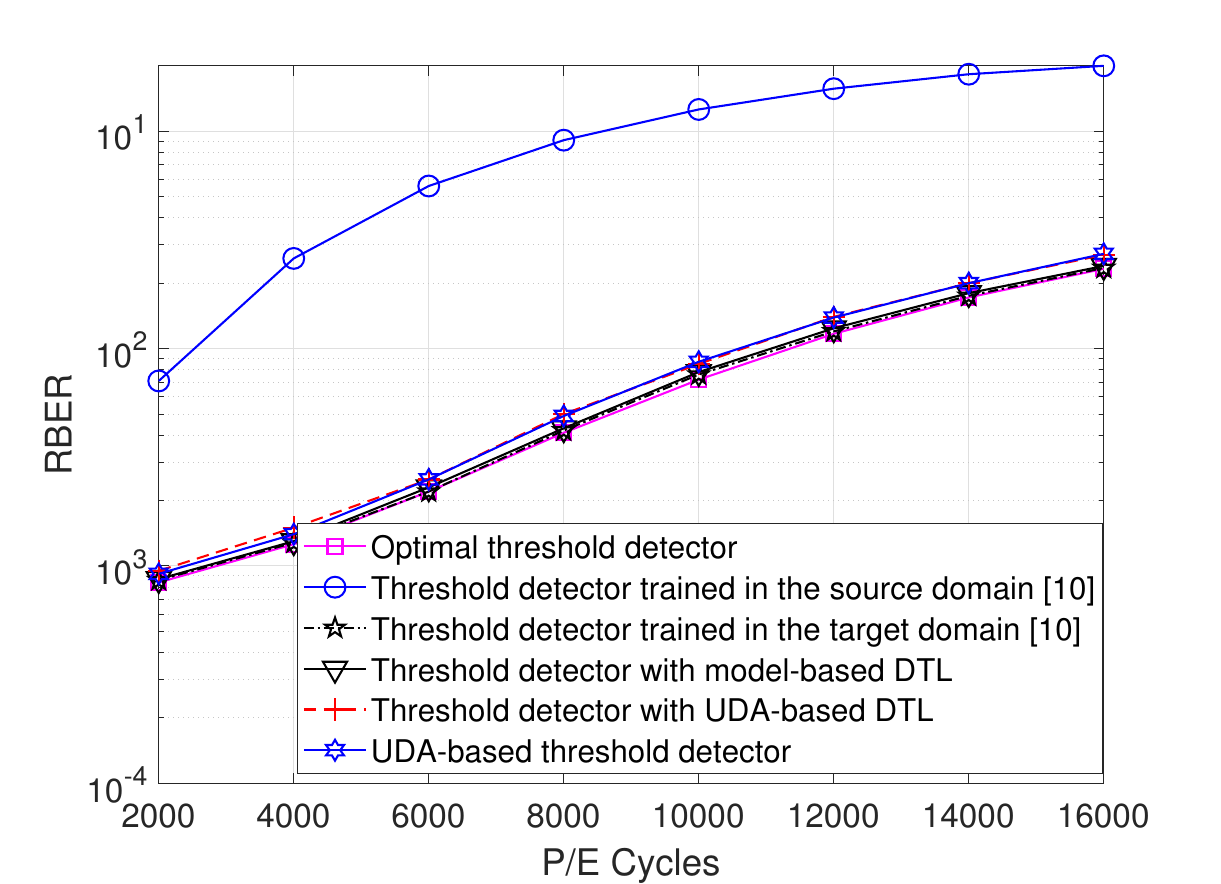}
	\caption{The RBER performance of different threshold detectors for MLC flash memory with Gamma distributed noises at $T^{\text{target}}=1.2\times 10^{4}$ hours.}
	\label{BER_Gamma}
\end{figure}

\subsection{Raw BER Performance}

Figs. \ref{BER_PE_Gaussian} and \ref{BER_time_Gaussian} illustrate the RBER performance of different RNNA threshold detectors for the MLC flash memory over different P/E cycles and retention time. The BER performance of the optimum threshold detector given by \eqref{BEP_2} is also included as a reference. First, we observe a large performance degradation of the RNNA threshold detector trained in the source domain, and it is due to the channel mismatch between the source domain and the target domain. This indicates the necessity of utilizing the knowledge of the target domain to improve the detection performance. Second, it can be observed that the RNNA threshold detector with the proposed model-based DTL achieves the performance of the RNNA threshold detector directly trained in the target domain. The RBERs of the two detectors also approach those of the optimum threshold detector. Third, performance of both the RNNA threshold detector with UDA-based DTL and the UDA-based threshold detector is slightly worse than the optimum performance. The reason is that only the mean of each voltage state is aligned between the source domain and the target domain in these detectors. As the variance of each voltage state is not aligned, there will be a threshold voltage PDF mismatch between the two domains.

The RBERs of different threshold detectors for the TLC flash memory is shown in Fig. \ref{BER_PE_Gaussian_TLC}. Different from the results for the MLC case, it can be seen that for the TLC flash memory, the performance of both the RNNA threshold detector with UDA-based DTL and the UDA-based threshold detector closely approaches the optimum performance. The reason for this interesting result is that the TLC flash memory channel is more ``symmetric" than the MLC flash memory channel. More specifically, the TLC flash memory cell has more threshold voltage states than the MLC one. Except for the erase state, the variances of all the programmed states are close to each other. This implies that the TLC flash memory has a higher percentage of voltage states with similar variances than the MLC flash memory. The proposed UDA is based on the mean of each voltage state, while the variance is not aligned. Hence, the proposed method is better for the TLC case, leading to better performance of the threshold detection schemes using UDA. 

Finally, to further verify the effectiveness of our proposed DTL approaches for different noise distributions, we consider the case that the noises in the target domain follow a Gamma distribution, while the noise PDF in the source domain remains Gaussian. It can be seen from Fig. \ref{BER_Gamma} that although the noise distributions of the source domain and target domain are different, our proposed DTL-based detectors can still achieve near-optimal performance. This indicates that the stacked RNN in Fig. 2 is capable of learning the detection of different noise distributions, and the proposed DTL approaches can effectively transfer the knowledge from the source domain to the target domain even if they have different noise PDFs. To conclude, Figs. \ref{BER_PE_Gaussian}-\ref{BER_Gamma} indicate that the proposed threshold detectors can effectively combat the unknown channel offset caused by the P/E cycling and data retention.

\begin{figure}[t]
\centering
\includegraphics[height=2.5in,width=3.5in]{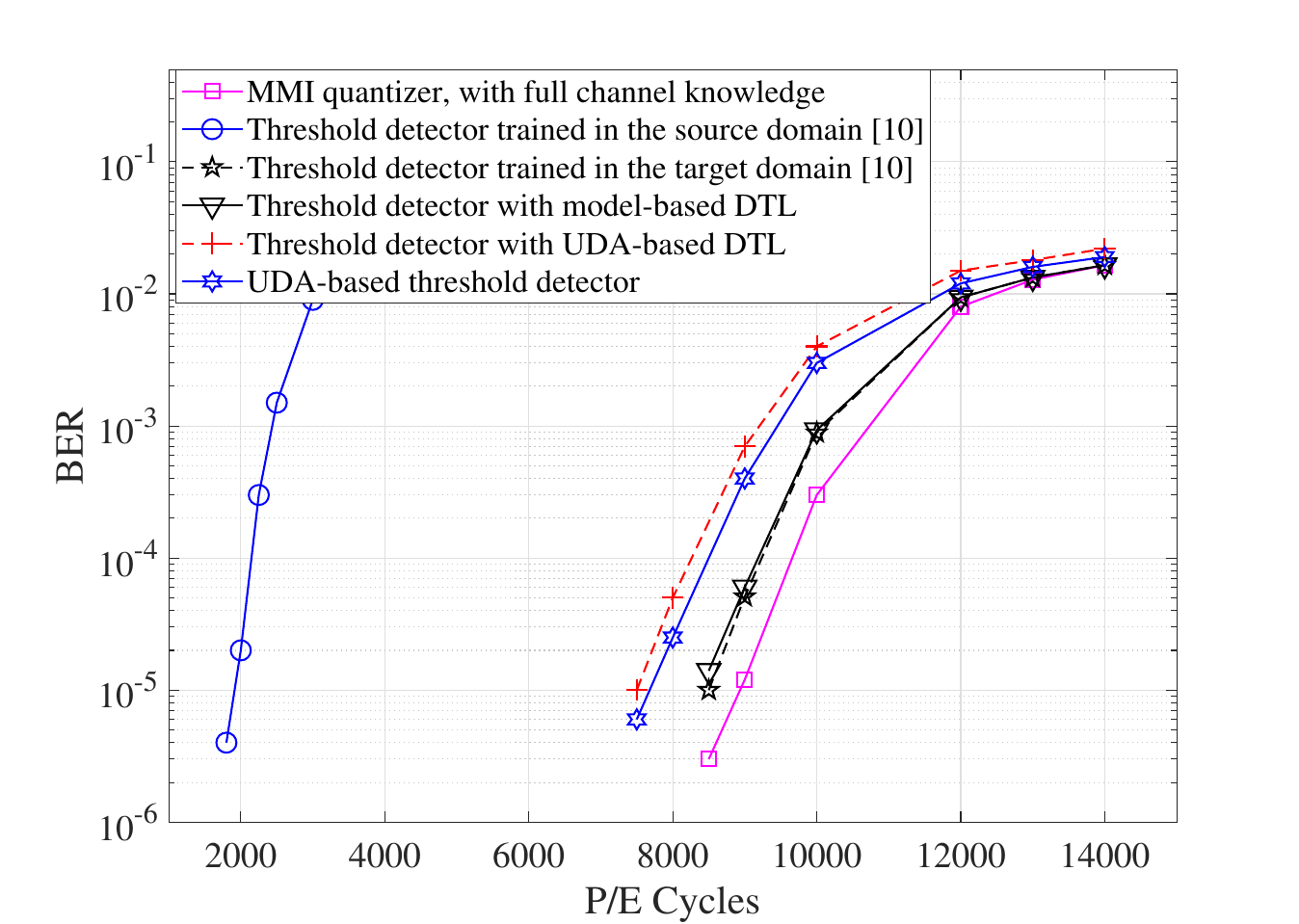}
\caption{The BER performance of the LDPC code with different threshold detectors for MLC flash memory at  $T^{\text{target}}=1.2\times 10^{4}$ hours.}
\label{BER_ldpc_PE_Gaussian}
\end{figure}

\begin{figure}[t]
\centering
\includegraphics[height=2.5in,width=3.4in]{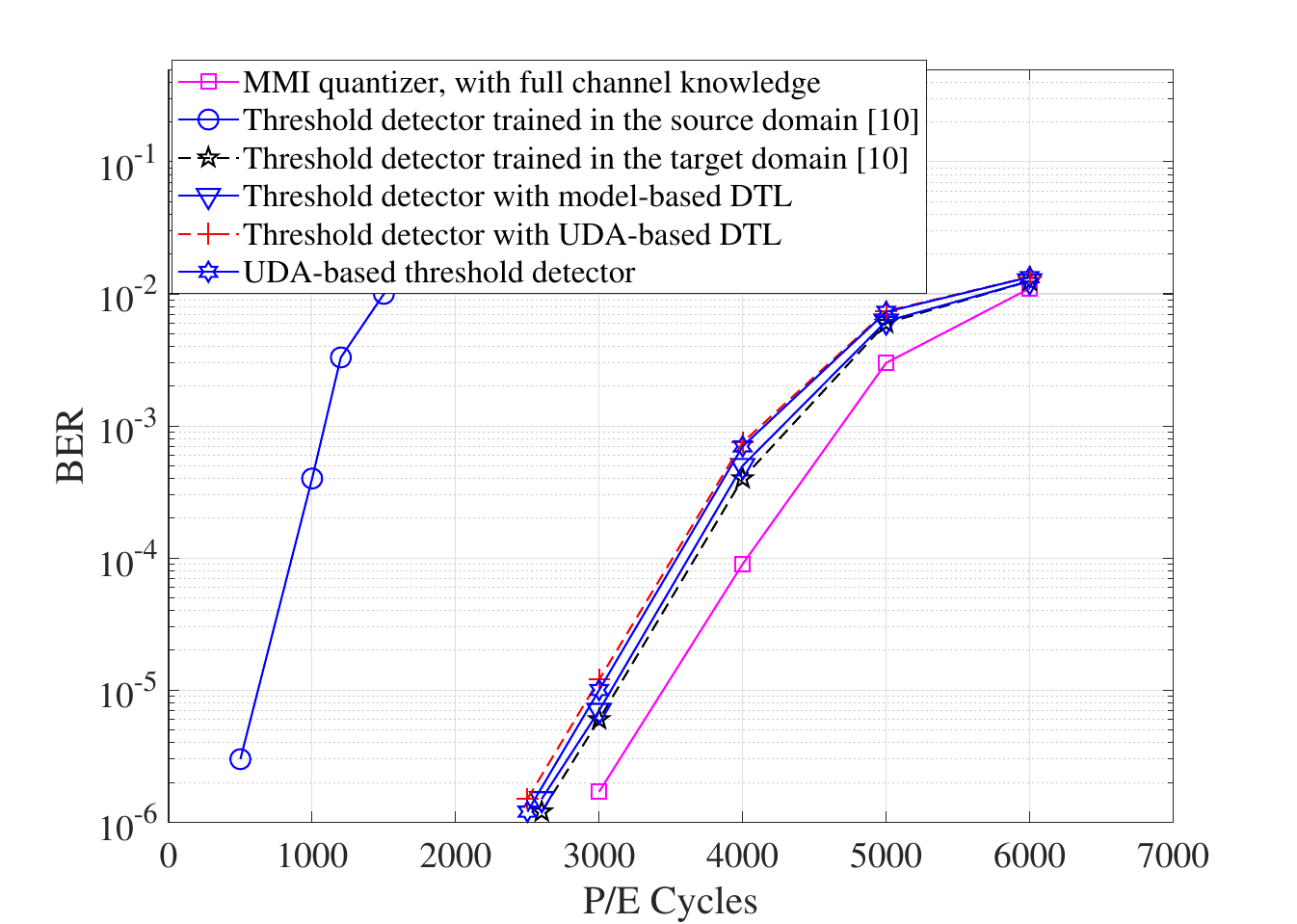}
\caption{The BER performance of the LDPC code with different threshold detectors for TLC flash memory at $T^{\text{target}}=1\times 10^{3}$ hours.}
\label{BER_ldpc_time_Gaussian}
\end{figure}

\begin{figure}[t]
	\centering
	\includegraphics[height=2.5in,width=3.5in]{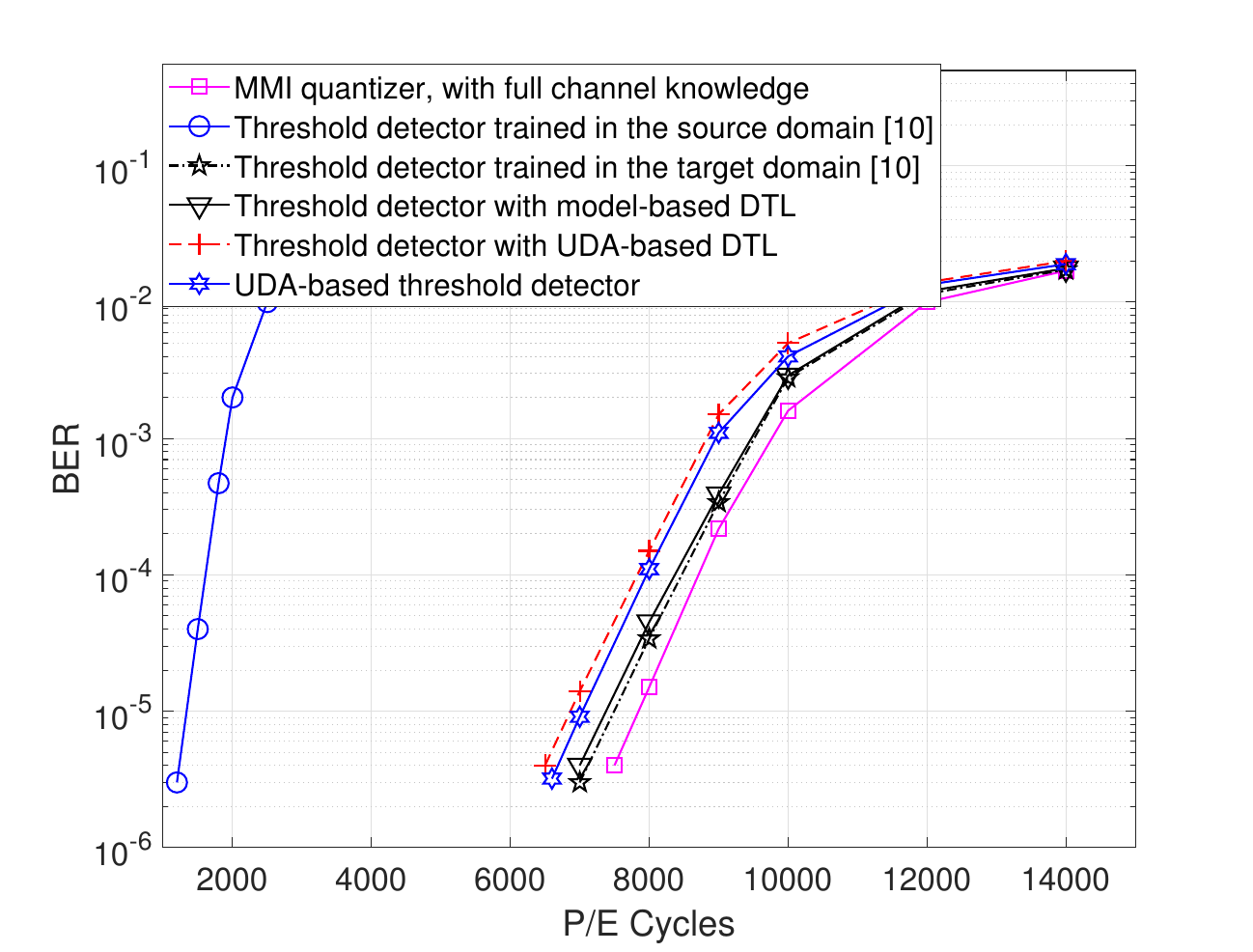}
	\caption{The BER performance of the LDPC code with different threshold detectors for MLC flash memory with Gamma distributed noises at  $T^{\text{target}}=1.2\times 10^{4}$ hours.}
	\label{ldpc_Gamma}
\end{figure}

\subsection{Coded BER Performance}

To examine the decoding performance of ECCs with learned thresholds, an irregular LDPC code with codeword length of 4544 bits and information length 4096 bits \cite{aslam2016read} is employed. The degree distribution of this code is given by
\begin{align} \nonumber
\lambda(x)&= 0.0682x+0.1822x^2+0.1329x^3+0.6167x^4,     \\
\rho(x)&= 0.22x^{38} + 0.78x^{39}.
\end{align}
This LDPC code is constructed by the progressive edge-growth algorithm \cite{hu2005regular}. The decoding algorithm is the normalized min-sum (NMS) algorithm \cite{chen2005reduced} with at most 20 decoding iterations. For the threshold detectors, the initial LLR is set to $5$ or $-5$, depending on the detected bit being $0$ or $1$, respectively. Furthermore, as a benchmark of the coded BER performance, we design the read thresholds by  MMI of the quantized flash memory channel \cite{wang2014enhanced}, which is based on the ideal assumption that the perfect channel knowledge is known. We use the corresponding BERs as the performance benchmark.
 

Figs. \ref{BER_ldpc_PE_Gaussian} and \ref{BER_ldpc_time_Gaussian} present the BER performance of the LDPC codes with the MMI quantizer and our proposed threshold detectors over different P/E cycles, for the MLC and the TLC flash memories, respectively. Similar to the RBER performance shown in Fig. \ref{BER_PE_Gaussian},  Fig. \ref{BER_ldpc_PE_Gaussian} shows that for the MLC case, the BER performance of the RNNA threshold detector trained in the source domain is much worse than that of other detectors. The LDPC-coded BER performance using the RNNA threshold detector with the model-based DTL can achieve that of the RNNA threshold detector that is directly trained in the target domain, with much less training data. The performance of the RNNA threshold detector with UDA-based DTL and the UDA-based threshold detector is slightly worse than that with model-based DTL, which is also consistent with our RBER performance shown in Fig. \ref{BER_PE_Gaussian}. Finally, it can be seen that the performance of our proposed threshold detectors is slightly worse than that of the MMI quantizer, due to the lack of channel PDF.

However, for the TLC flash memory, Fig. \ref{BER_ldpc_time_Gaussian} shows that the performance of both the RNNA threshold detector with UDA-based DTL and the UDA-based threshold detector can almost achieve that of the RNNA threshold detector that is directly trained in the target domain. This is again consistent with the uncoded case illustrated by Fig. \ref{BER_PE_Gaussian_TLC}.  Furthermore, the LDPC-coded BER performance under non-Gaussian Gamma distributed noises in the target domain is illustrated in Fig.  \ref{ldpc_Gamma}. It shows that all the proposed detectors can achieve near-optimal performance, and the UDA-based detectors perform slightly worse than the threshold detector with model-based DTL.   Similar to Fig. \ref{BER_Gamma}, the LDPC-coded BER performance indicates that our proposed DTL-based detectors have the ability to transfer the knowledge to different noise distributions.

The simulation results illustrated by Figs. \ref{BER_PE_Gaussian}-\ref{ldpc_Gamma} demonstrate that our proposed threshold detectors based on the DTL and UDA can  almost achieve the performance of the RNNA threshold detector that is directly trained in the target domain, with much less training data. This is because the proposed detection schemes have transferred the knowledge from the source domain to the target domain through model parameters and domain adaptation, which can achieve better error rate performance with much less training complexity.

\section{Conclusion}
In this paper, we have formulated the data detection for the flash memory channel as a TL problem, and proposed a model-based DTL algorithm to effectively reduce the number of training samples and labels. We have further proposed a UDA-based DTL algorithm to cope with the scenarios where the channel has a large unknown offset such that the labels cannot be obtained. Furthermore, we have derived the RNNA threshold detectors with the proposed DTL algorithms.  A UDA-based threshold detection scheme has also been proposed which completely avoids the use of neural network. Both the channel raw error rate analysis and simulation results demonstrate that our proposed DTL-based detection schemes can achieve the near-optimal BER performance with much less training data and/or without using any labels. A possible future research topic is the design and optimization of the multi-bit channel quantizer for the proposed DTL-based detectors.

\small
\bibliographystyle{IEEEtran}
\bibliography{postdoc_refs}

\end{document}